\documentstyle[12pt,epsf]{article}
\newcommand{\Appendix}[1]%
    {%
     \section{#1}%
      }

\newcommand \bra[1]{\left< {#1} \,\right\vert}
\newcommand \ket[1]{\left\vert\, {#1} \, \right>}

\newcommand{\simgt}{\hbox{ \raise3pt\hbox to 0pt{$>$}\raise-3pt\hbox{$\sim$} }}
\newcommand{\simlt}{\hbox{ \raise3pt\hbox to 0pt{$<$}\raise-3pt\hbox{$\sim$} }}

\newcommand{\nn}{\nonumber}
\newcommand{\be}{\begin{equation}}
\newcommand{\ee}{\end{equation}}
\newcommand{\bea}{\begin{eqnarray}}
\newcommand{\eea}{\end{eqnarray}}

\def\lQ{\Lambda_{\rm QCD}}

\def\als{\alpha_{\rm s}}
\def\siml{{\ \lower-1.2pt\vbox{\hbox{\rlap{$<$}\lower6pt\vbox{\hbox{$\sim$}}}}\ }}

\def\mbar{\overline{m}}

\def\msbar{{\overline{\rm MS}}}

\def\to{\rightarrow}

\begin{document}
\begin{titlepage}
\begin{flushright}
\tt
CERN-TH/2001-201 \\ 
IFUM--687--FT  \\ 
TU-625
\end{flushright}
\vspace*{2cm}
\begin{center}
\begin{Large}
{\bf Quarkonium spectroscopy and perturbative QCD:\\
\vspace{4mm}  massive quark-loop effects}
\\[17mm]
\end{Large}
  {\large
    N.~Brambilla$^1$, Y.~Sumino$^2$ and A.~Vairo$^3$
    }
  \\[8mm]
  {\it
    $^1$
    INFN and Dipartimento di Fisica dell'Universit\`a di Milano  \\
    via Celoria 16, 20133 Milan, Italy
    }
  \\[8mm]
  {\it
    $^2$
    Department of Physics, Tohoku University\\
    Sendai, 980-8578 Japan
    }
  \\[8mm]
  {\it
    $^3$
    Theory Division CERN, 1211 Geneva 23, Switzerland
    }
\end{center}
\vspace{1cm}
\begin{abstract}
\baselineskip=16 pt 
\noindent 
We study the spectra of the bottomonium and $B_c$ states within 
perturbative QCD up to order $\als^4$. The ${\cal O}(\lQ)$ renormalon
cancellation between the static potential and the pole mass is performed 
in the $\varepsilon$-expansion scheme. We extend our previous analysis by
including the (dominant) effects of non-zero charm-quark mass in loops up to the 
next-to-leading non-vanishing order $\varepsilon^3$. We fix the $b$-quark $\msbar$ mass 
$\overline{m}_b \equiv m_b^{\overline{\rm MS}}(m_b^{\overline{\rm MS}})$ 
on $\Upsilon(1S)$ and compute the higher levels.
The effect of the charm mass decreases $\overline{m}_b$ by about 11 MeV 
and increases the $n=2$ and $n=3$ levels by about 70--100 MeV and 
240--280 MeV, respectively. We provide an extensive quantitative analysis. 
The size of non-perturbative and higher order contributions is discussed by comparing 
the obtained predictions with the experimental data. An agreement of the 
perturbative predictions and the experimental data depends crucially on the 
precise value (inside the present error) of $\als (M_Z)$. We obtain 
$m_b^{\overline{\rm MS}}(m_b^{\overline{\rm MS}}) = 4190 \pm 20 \pm 25 \pm 3 ~ {\rm MeV}$.
\end{abstract}
\end{titlepage}
\vfill\eject

\setcounter{footnote}{0}
\pagenumbering{arabic}

\section{Introduction}
\label{s1}
Traditional methods for theoretical investigations of heavy-quarkonium 
spectra have been based on various phenomenological potential models in the frame 
of non-relativistic quantum mechanics. These phenomenological potentials 
were tuned to reproduce the observed quarkonium spectra (and some other physical observables) 
and have been (in most cases) successful in predicting physical 
observables of the quarkonia, such as leptonic widths and transition rates among 
different levels, besides reproducing the energy levels. They have elucidated various 
properties of the bound states and essentially established, from consistency, 
the non-relativistic nature of the quarks inside quarkonia. 
For some reviews we refer to \cite{gro}. An apparent deficit 
of these approaches is, however, the difficulty in relating the phenomenological parameters 
to the fundamental parameters of QCD.

Recently we have reported in \cite{bsv1} new results on the spectroscopy of heavy
quarkonia (charmonium, bottomonium and $B_c$) computed in perturbative QCD 
up to order $\als^4$, once the ${\cal O}(\lQ)$ renormalon cancellation between the 
static potential and the pole mass \cite{renormalon1,renormalon2} has been implemented. 
In order to realize this we have adopted the so-called $\varepsilon$-expansion 
scheme \cite{hlm}. (In the present work we will use the same scheme.)
The major results of that analysis have been as follows.
(1) Once the cancellation of the leading renormalon has been incorporated,
the perturbative series turns out to be convergent and to reproduce reasonably well 
the gross structure of the bottomonium spectrum at least up to some of the 
$n=3$ levels. (2) The constraints on non-perturbative and higher-order contributions 
to the bottomonium spectrum, set by the comparison of the calculation with 
the experimental data, indicate that these are smaller than usually believed. 

In this paper we improve the analysis of \cite{bsv1} by including the effects of 
the non-zero charm-quark mass loops on the spectra of bottomonium and $B_c$. 
The typical scales of the bottomonium states as well as of the $B_c$ 
(as obtained in \cite{bsv1}) are either close to or less than the charm-quark mass. 
Hence, charm-mass effects are expected and actually turn out to be numerically 
not negligible. These effects have been computed for the pole-mass--$\msbar$-mass 
relation at leading (non-vanishing) order in \cite{gray} and at the next-to-leading 
order in the limit of small charm mass in \cite{hoangcm}. 
In the binding-energy expansion they have been calculated at leading 
(non-vanishing) order in \cite{eiras} and at the next-to-leading order, only 
for the $\Upsilon(1S)$ mass, in \cite{hoangcm}. 
In the present work we implement and discuss the above results and 
derive some new formulas valid for the inclusion of charm-mass effects 
at the next-to-leading order for all quarkonium excited states.  
This will allow us to fully take into account charm-mass effects 
in the spectra of bottomonium and $B_c$ up to order $\varepsilon^3$ 
in the $\varepsilon$-expansion. Finally, we re-examine whether, by including 
these effects, the above conclusions (1) and (2) of \cite{bsv1} still hold.

The paper is organized as follows. In Sec. \ref{s2} we review the theoretical 
framework and the results of Ref. \cite{bsv1}. In Sec. \ref{seccharm} we present 
all the formulas needed for the inclusion of the non-zero charm-mass effects 
up to order $\varepsilon^3$ in the bottomonium and $B_c$ energy levels.  
In Sec. \ref{san} we perform the numerical analyses on the bottomonium 
and $B_c$ spectra that include the above effects. 
In Sec. \ref{scd} we give some conclusions.

\section{Quarkonium Spectrum with Massless Quark Loops}
\label{s2}
The heavy-quarkonium dynamics is characterized by at least two dynamically
generated scales: the (soft) scale $mv$, of the order of the momentum transfer and 
the (ultrasoft) scale $mv^2$ of the order of the energy of the quark in the 
bound state ($v$ being the typical size of the heavy-quark velocity in a quarkonium state).
The magnitude and nature of the non-perturbative corrections depend 
on the relative size of $\lQ$ with respect to the dynamically generated scales. 
Different scenarios are then possible. These have been systematically 
studied in \cite{long}, where the effective field theories corresponding to them, 
generically denoted as potential non-relativistic QCD \cite{ps}, have been constructed. 
In \cite{bsv1} we have discussed our work in relation to these scenarios. 
That discussion applies also here. In order to facilitate the reading,  
let us summarize the main points. We assume to be in a situation where $mv \gg \lQ$ 
and the system to be mainly Coulombic. In this scenario non-perturbative corrections 
may be carried by local or non-local gluon condensates. Due to the fact that, in both cases, 
a direct numerical evaluation turns out to be quite uncertain, we chose not to correct 
explicitly our perturbative formulas with them (as done, for instance, in \cite{yn3,py}), 
but rather to infer their size, as well as that of the neglected higher-order 
perturbative contributions from the eventual disagreement (and uncertainties) 
of the perturbative results from the experimental data. Among the neglected higher-order 
perturbative contributions we also include the $\als^5\ln \als$ contributions, 
which are known from \cite{n3lo}, but whose dominance over other ${\cal O}(\als^5)$ 
corrections is unclear. Therefore, using the effective field theory language of \cite{long},  
we will consider in the following heavy quarkonium as described by potential 
non-relativistic QCD in the perturbative regime up to order $\als^4$.
This will correspond to just considering the quarkonium spectrum up to 
that order as obtained from standard perturbative QCD.

\subsection{Perturbative expansions}
In this section, in order to set up some basic formulas, we review the 
energy (mass) of a quarkonium state $X$ made by two heavy quarks in the theory 
with $n_l$ massless quarks only. Namely, we do not take into account, for the
moment, the effects of non-zero masses in the light-quark loops, neither in 
the binding energy nor in the pole-mass expansion. These will be considered 
in Sec. \ref{seccharm}. 

We express the quarkonium mass as a series expansion in the $\overline{\rm MS}$ 
coupling constant $\als^{(n_l)}(\mu)$ defined in the theory with $n_l$ 
massless quarks only. Consider the quantum-mechanical Hamiltonian
\bea
{H} =  m_{1,{\rm pole}}+m_{2,{\rm pole}} +
{H}_0 + {H}_1 + {H}_2 + \cdots ,
\eea
where\footnote{
The term $H_2$, in its unitary-equivalent form, can be read 
from Ref. \cite{yn3} supplemented with the two-loop static potential 
calculated in \cite{Peter}.}
\bea
\rule[-6mm]{0mm}{6mm}
{H}_0 &=& \frac{\vec{p}\,^2}{2m_r} \,
{- \, \frac{4\, \als^{(n_l)}(\mu)}{3\, r}} , 
\label{h0}
\\ \rule[-8mm]{0mm}{6mm}
{H}_1 &=& 
{- \, \frac{4\, \als^{(n_l)}(\mu)}{3\, r} \,
\biggl( \frac{\als^{(n_l)}(\mu)}{4\pi} \biggr) \,
\biggl\{ 2 \beta^{(n_l)}_0 \, \ell } + a^{(n_l)}_1 \biggr\}
, 
\\
{H}_2 &=& 
{- \, \frac{4\, \als^{(n_l)}(\mu)}{3\, r}\,
\biggl( \frac{\als^{(n_l)}(\mu)}{4\pi} \biggr)^2 \,
\biggl\{ \beta^{(n_l)\, 2}_0 \, \Bigl( 4\ell^2 + \frac{\pi^2}{3} \Bigr) 
}
+ 2 (\beta^{(n_l)}_1+2\beta^{(n_l)}_0 a^{(n_l)}_1)\ell + a^{(n_l)}_2
\biggr\}
\nonumber \\ &&
- \frac{(1+3x) \, \vec{p}\,^4}{32\, m_r^3} \,
+ \frac{\pi (1+x) \, \als^{(n_l)}(\mu)}{3\, m_r^2} \, \delta^3(\vec{r})
- \frac{(1-x)\, \als^{(n_l)}(\mu)}{6\, m_r^2 \, r} \biggl(
\vec{p}\,^2 + \frac{1}{r^2} r_i r_j p_j p_i \biggr)
\nonumber \\ &&
- \frac{(1-x)\, \als^{(n_l)}(\mu)}{6 \, m_r^2}
\biggl\{ \frac{S^2}{r^3} - 3 \frac{(\vec{S}\cdot\vec{r})^2}{r^5}
- \frac{4\pi}{3}(2S^2-3) \delta^3(\vec{r}) 
\biggr\} 
- \frac{ \als^{(n_l)}(\mu)^2}{m_r \, r^2}
\nonumber
\\ &&
+\frac{ (3-x) \, \als^{(n_l)}(\mu)}{6\, m_r^2 \, r^3} \, \vec{L}\cdot\vec{S}
+ \frac{\als^{(n_l)}(\mu)}{3 \, r^3}
\left(
\frac{1}{{m_{1,{\rm pole}}}^2} - \frac{1}{{m_{2,{\rm pole}}}^2} 
\right)
\vec{L}\cdot (\vec{S}_1 - \vec{S}_2)
. \label{h2}
\eea
We have defined the reduced pole mass as 
$m_r = m_{1,{\rm pole}}m_{2,{\rm pole}}/(m_{1,{\rm pole}}+m_{2,{\rm pole}})$ 
and $x = 1-4m_r/(m_{1,{\rm pole}}+m_{2,{\rm pole}})$; when the two masses are equal, $x = 0$. 
The $\beta^{(n_l)}_k$'s denote the coefficients of the QCD beta function: 
\bea
\beta^{(n_l)}_0 = 11 - \frac{2}{3} n_l, 
~~~~~~
\beta^{(n_l)}_1 = 102 - \frac{38}{3} n_l,
\eea
$\ell$, $a^{(n_l)}_1$ and $a^{(n_l)}_2$ are given by
\bea
&&
\ell = \ln (\mu r) + \gamma_E ,
\\ &&
a^{(n_l)}_1 = \frac{31}{3} - \frac{10}{9} n_l ,
\\ &&
a^{(n_l)}_2 = {\frac{4343}{18}} + 36\,{{\pi }^2} +   66\,{\zeta_3} - 
  {\frac{9\,{{\pi }^4}}{4}} - 
   \left( {\frac{1229}{27}} + {\frac{52\,{\zeta_3}}{3}} \right)\,{n_l}  
+ {\frac{100}{81}} \,{n_l^2} .
\eea

Up to ${\cal O}(\als^4 m)$, the energy of a heavy quarkonium state $X$, 
identified by the quantum numbers $n,l,s$ and $j$, is derived from 
the perturbative expansions of the energy eigenvalues\footnote{
The full formula up to ${\cal O}(\als^4 m)$ for the $S$ state spectrum was derived 
in \cite{py} and later confirmed in \cite{my}; additional corrections necessary 
for the spectrum of $l \geq 1$ states can be found in \cite{yn3} and 
the formula for the unequal mass and $l=0$ case in \cite{bcbv}.}
of the above Hamiltonian:
\bea
&&
E_{X}(\mu,\als^{(n_l)}(\mu),m_{i,{\rm pole}}) =  m_{1,{\rm pole}}+m_{2,{\rm pole}} 
+ E^{n_l}_{{\rm bin},X}(\mu,\als^{(n_l)}(\mu),m_{i,{\rm pole}}) ,
\label{spectrum}
\\ &&
E^{n_l}_{{\rm bin},X}(\mu,\als^{(n_l)}(\mu),m_{i,{\rm pole}}) = 
- \, \frac{8}{9n^2} \, \als^{(n_l)}(\mu)^2  m_{r}
\sum_{k=0}^2 \varepsilon^{k+1} \left( \frac{\als^{(n_l)}(\mu)}{\pi} \right)^k P_k (L_{nl}),
\label{spectrum2}
\eea
where $\varepsilon=1$ is the parameter that will be used in order to properly organize 
the perturbative expansion in view of the ${\cal O}(\lQ)$ renormalon cancellation \cite{hlm}. 
$P_k(L_{nl})$ is a $k^{\rm th}$-degree polynomial of $\displaystyle L_{nl} 
\equiv \ln [\, 3n\mu / (8 \als^{(n_l)}(\mu)m_{r}) \, ] +S_1(n+l)+\frac{5}{6}$, and
the harmonic sums are defined as $\displaystyle S_p (q) \equiv \sum_{k=1}^{q} \frac{1}{k^p}$.
It is convenient to decompose the polynomials into renormalization-group
invariant subsets:
\bea
P_0 &=& 1,
\\
P_1 &=& \beta^{(n_l)}_0  \, L_{nl} + c_1 ,
\\
P_2 &=&
\frac{3}{4} \beta^{(n_l)\,2}_0\, {L_{nl}^2} + 
  \left( - \frac{1}{2} \beta^{(n_l)\,2}_0 
             + 
     {\frac{1}{4}\beta^{(n_l)}_1 } + 
     {\frac{3}{2} \beta^{(n_l)}_0 {c_1}} \right) L_{nl}  + 
  {c_2^{(n_l)}} .
\eea
$c_1$ and $c_2^{(n_l)}$ are given by\footnote{
This corrects the formula given for $c_2^{(n_l)}$ in \cite{bsv1} in the
situation when $m_{1,{\rm pole}} \neq m_{2,{\rm pole}}$ and $l>0$ hold at the same time.
Since, however, all the numerical analyses of \cite{bsv1} correspond to the equal-mass case
or to $l=0$ states, they remain valid.
}
\bea
c_1~~
&=& -4 ,
\\
c^{(n_l)}_2 &=&
- 
  {\frac{16\,{{\pi }^2}\,\left\{ 2s ( s+1 ) 
         ( 1 - x )  + 3 x \right\} }{27\,n}} 
\,     \delta_{l0}
+ 
  {\frac{8\,{{\pi }^2}\, \lambda (l,s,j)  }{9\,n\,l\,
      \left( l+1 \right) \,\left( 2\,l +1 \right) }} 
\,
      \left( 1 \! - \! \delta_{l0} \right)
+ 
\beta^{(n_l)\,2}_0 \,\nu (n,l) 
\nonumber \\&&
- 
  {\frac{\left( 11 + x \right) \,{{\pi }^2} }{9\,{n^2}}} 
+ 
  {\frac{68\,{{\pi }^2}}{9\,n\,\left( 2\,l +1\right) }}
+
{\frac{473}{16}} + {\frac{9\,{{\pi }^2}}{2}} 
+ {\frac{33\,\zeta_3}{4}} 
- 
  {\frac{9\,{{\pi }^4}}{32}} 
- 
  {n_l}\,\left( {\frac{109}{72}} +
     {\frac{13\,\zeta_3}{6}} \right) ,
\nonumber \\ &&
\eea
where $\lambda (l,s,j)$ represents the fine and hyperfine splittings for $l>0$.
If $j = l \pm 1$ or $m_{1,{\rm pole}} = m_{2,{\rm pole}}$ ($x=0$), it is given by
\bea
\lambda (l,s,j) = - ( 1 - x ) D_S - ( 3 - x ) X_{LS},
\eea
with
\bea
&&
D_{S} \equiv
\left< 
3 \frac{(\vec{r}\cdot \vec{S})^2}{r^2} - \vec{S}^2 
\right>
=
\frac{
2 l (l+1) s (s+1) - 3 X_{LS} - 6 X_{LS}^2
}{
(2l-1)(2l+3)
},
\\&&
X_{LS} \equiv
\left< \vec{L}\cdot \vec{S} \right>
= \frac{1}{2}\,
\left[ j(j+1)-l(l+1)-s(s+1) \right] .
\eea
If $j=l$ and $m_{1,{\rm pole}} \neq m_{2,{\rm pole}}$ ($x>0$), the last term of
Eq. (\ref{h2}) induces a mixing between the $s=0$ and $s=1$ states, and the total
spin $s$ is no longer a good quantum number (see, for instance, \cite{kiselev}).
In this case, the splitting is given by\footnote{
The corresponding energy eigenstate reads
$\ket{\pm} = 
[ \, b \, \ket{s=0} - \lambda_{\pm} \ket{s=1} \, ] / {\sqrt{|b|^2+\lambda_{\pm}^2}}$,
where $b = 2 \sqrt{x\, l\, (l+1)}$ up to a convention-dependent phase.
}
\bea
\lambda (l, \pm , j=l ) = 1 \pm \sqrt{ 1 + 4\, x\, l\, (l+1) } .
\eea
The term $\nu (n,l)$ is given by\footnote{
The infinite sum can easily be evaluated analytically 
in terms of $\zeta_3$, etc.  for given values of $n$ and $l$, e.g. by using {\it Mathematica}.}
\bea
&&
\nu (n,l)=
\frac{\pi^2}{8}
- \frac{1}{2} \, S_2(n+l)
+ \frac{n}{2} \frac{(n+l)!}{(n-l-1)!}
\sum_{k=1}^{\infty}
\frac{(n-l+k-1)!}{(n+l+k)! \, k^3}
\nonumber \\&& 
~~~~~~~~~~~~
+ \frac{(n-l-1)!}{2(n+l)!}
\sum_{k=1}^{n-l-1}
\frac{(2l+k)! \, (2k+2l-n)}
{(k-1)! \, (k+l-n)^3} .
\label{nu}
\eea
It is understood that the last term is zero if $n-l<2$.

Next we rewrite the series expansion of $E_{X}$ in terms of the $\overline{\rm MS}$ masses. 
This is done by expressing the pole masses $m_{i,{\rm pole}}$ in terms of the  
renormalization-group-invariant $\overline{\rm MS}$ masses 
$\displaystyle \overline{m}_i \equiv m_{i,\overline{\rm MS}} (m_{i,\overline{\rm MS}})$.
In the theory with $n_l$ massless quarks only, we have up to ${\cal O}(\als^3)$:
\begin{eqnarray}
m_{i,{\rm pole}} = \overline{m}_i \left\{ 1 + {4\over 3} \varepsilon {\als^{(n_l)}(\overline{m}_i)\over \pi} 
+ \varepsilon^2  \left({\als^{(n_l)}(\overline{m}_i)\over \pi}\right)^2 d^{(n_l)}_1 
+ \varepsilon^3 \left({\als^{(n_l)}(\overline{m}_i)\over \pi}\right)^3 d^{(n_l)}_2 \right\}.
\label{pole}
\end{eqnarray}
The coefficients $d^{(n_l)}_1$ and $d^{(n_l)}_2$ are given in \ref{appole}.

Note that the counting in $\varepsilon$ in Eq. (\ref{spectrum2}) and Eq. (\ref{pole}) 
does not reflect the order in $\als$ but the wanted renormalon cancellation \cite{hlm}.
One way to understand this is to consider that in the sum of the pole quark masses and the 
static QCD potential, $\sum_i m_{i,{\rm pole}} + V_{\rm QCD}(r)$, 
the renormalon cancellation takes place without reordering of power counting 
in $\als$ \cite{renormalon1,renormalon2}. 
The extra power of $\als$ comes in the energy-level expansion when the dynamical variable $r^{-1}$ 
is replaced by the dynamical scale $\bra{nlsj} r^{-1} \ket{nlsj} \sim C_F \als
m_r / n$, where $C_F=4/3$. 
Moreover, in order to realize the renormalon cancellation at each order of the
expansion, it is necessary to expand $m_{i,{\rm pole}}$ and $E^{n_l}_{{\rm bin},X}$ 
in the same coupling \cite{mar,hlm,sumino}; we therefore 
express $\als^{(n_l)}(\overline{m}_i)$ in (\ref{pole}) in terms of $\als^{(n_l)}(\mu)$: 
\bea
\als^{(n_l)}(\overline{m}_i)&=& \als^{(n_l)}(\mu) \left\{ 
1 + \varepsilon \, \frac{\als^{(n_l)}(\mu)}{\pi}  \,\frac{\beta^{(n_l)}_0}{2} 
\ln\left(\frac{\mu}{\overline{m}_i}\right) \right. \nn\\
& &\qquad\qquad\quad   \left.
+ \varepsilon^2 \left( \frac{\als^{(n_l)}(\mu)}{\pi} \right)^2 \,
\Biggl[ \frac{\beta^{(n_l)\,2}_0}{4} \, \ln^2 \left(\frac{\mu}{\overline{m}_i}\right)
+ \frac{\beta^{(n_l)}_1}{8} \ln\left(\frac{\mu}{\overline{m}_i}\right) \Biggr] \right\}.
\label{alphams}
\eea

Inserting Eqs. (\ref{alphams}) and (\ref{pole}) into 
Eqs. (\ref{spectrum2}) and (\ref{spectrum}), we get an expression for the
energy levels of the heavy quarkonium states, which depends on $\mu$,
$\als^{(n_l)}(\mu)$ and $\overline{m}_i$, that we can organize as an expansion
in $\varepsilon$ up to order $\varepsilon^3$:
\begin{eqnarray}
& & E_{X}(\mu, \als^{(n_l)}(\mu),\overline{m}_i) \nn\\
& & 
= \left[ \sum_{i=1}^2 
\overline{m}_i \left\{ 1 + {4\over 3} \varepsilon {\als^{(n_l)}(\overline{m}_i)\over \pi} 
+ \varepsilon^2  \left({\als^{(n_l)}(\overline{m}_i)\over \pi}\right)^2 d^{(n_l)}_1 
+ \varepsilon^3 \left({\als^{(n_l)}(\overline{m}_i)\over \pi}\right)^3 d^{(n_l)}_2 \right\} \right. \nn\\
& & \hspace{5mm}
\left.
+ E^{n_l}_{{\rm bin},X} \left(\mu,\als^{(n_l)}(\mu),
\overline{m}_i \left\{ 1 + {4\over 3} \varepsilon {\als^{(n_l)}(\overline{m}_i)\over \pi} 
+ \varepsilon^2  \left({\als^{(n_l)}(\overline{m}_i)\over \pi}\right)^2 d^{(n_l)}_1 \right\} \right) 
\right]_{\als(\mbar_i) = \hbox{\footnotesize Eq. (\ref{alphams})}} \label{en2}\\
\nn\\
& & 
\equiv \overline{m}_1 + \overline{m}_2 
+ E_{X}^{n_l\,(1)}(\mu, \als^{(n_l)}(\mu),\overline{m}_i) \varepsilon 
+ E_{X}^{n_l\,(2)}(\mu, \als^{(n_l)}(\mu),\overline{m}_i) \varepsilon^2 \nn\\
& & \hspace{5mm}
+ E_{X}^{n_l\,(3)}(\mu, \als^{(n_l)}(\mu),\overline{m}_i) \varepsilon^3 + \dots
\label{en2a}
\end{eqnarray}
Since the counting in $\varepsilon$ explicitly realizes the order $\lQ$ renormalon cancellation, 
and since $\als$ and $\overline{m}_i$ are short-range quantities, 
the obtained perturbative expansion (\ref{en2a}) is expected to show a better 
convergence than Eq. (\ref{spectrum2}).

\subsection{Physical parameters}
\label{parameter}
The input value for $\als$ that we will use is \cite{pdg}:
\be
\als^{(5)}(M_Z) = 0.1181\pm 0.0020. \label{apdg}
\ee
Throughout this paper, we evolve the coupling and match it to the couplings 
of the theory with $n_l=4$ and 3 successively, by solving the
renormalization-group equation perturbatively 
(analytically) at 4 loops (Eqs. (3) and (11) 
of Ref. \cite{running}).\footnote{
We take the matching scales as $\overline{m}_b$ and $\overline{m}_c$, respectively.}
We obtain 
\bea
\Lambda_{\msbar}^{(5)} &=& 210^{+24}_{-23} \> {\rm MeV}, \label{l5}\\ 
\Lambda_{\msbar}^{(4)} &=& 292^{+30}_{-28} \> {\rm MeV}, \label{l4}\\ 
\Lambda_{\msbar}^{(3)} &=& 333^{+31}_{-28} \> {\rm MeV}. \label{l3}
\eea

In Ref. \cite{bsv1} we calculated the bottomonium spectrum in the
$m_c = 0$ case, i.e. directly from the above Eq. (\ref{en2a}).\footnote{
In Ref. \cite{bsv1} major numerical results have been presented using the
coupling $\als^{(n_l)}(\mu)$ obtained by solving the 4-loop
renormalization-group equation {\it numerically}.
Here, we present the results computed using the above {\it analytic}
coupling (corresponding to Table 2 column (i) of \cite{bsv1}).
}  
In particular, by fitting the theoretical values of the masses of the $\Upsilon(1S)$ and $J/\psi$ 
with the experimental ones, $\overline{m}_b$ and $\overline{m}_c$ were found to be 
\bea
\overline{m}_b &=& 4201^{-19}_{+18} ~ {\rm MeV} 
\quad \hbox{(without charm-mass effects)},
\label{bmass0}
\\
\overline{m}_c &=& 1237^{-16}_{+16} ~ {\rm MeV},
\label{cmass0}
\eea
where we quote only the errors due to the uncertainty in $\als^{(5)}(M_Z)$ 
given in Eq. (\ref{apdg}).
We refer to \cite{bsv1} for an extended analysis of the other sources of uncertainty in the 
above determination. In \cite{bsv1}, with these values of the quark masses 
as input, other bottomonium masses have been calculated. 
The masses of the states $2S_1$, $1P_0$, $1P_1$, $1P_2$, $2P_0$ and $3S_1$ 
could be determined in a reliable way in the sense of \cite{bsv1} (see also Sec. \ref{san}).

\begin{figure}[htb]
\makebox[0truecm]{\phantom b}
\put(50,0){\epsfxsize=12.0truecm\epsffile{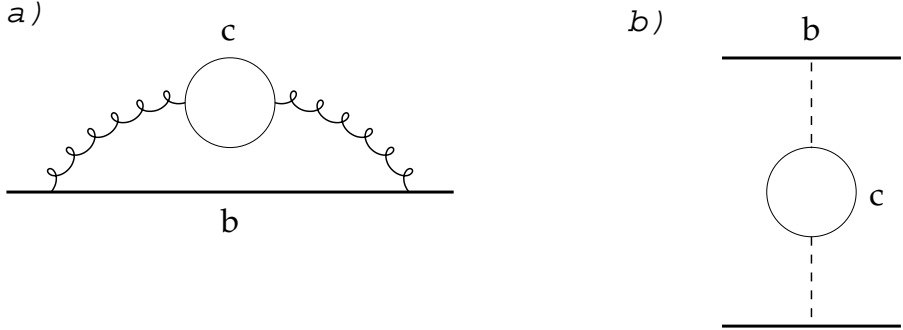}}
\vspace{5mm}
\caption{\footnotesize \it Feynman graphs responsible for the leading charm-mass effects to the pole mass 
a) and to the Coulomb potential b).}
\label{fig1loop}
\end{figure}

\section{Charm-Mass Effects}
\label{seccharm}
In this section we discuss how Eqs. (\ref{spectrum}) and (\ref{pole}) get modified 
in the bottomonium case by including finite charm-mass effects up to order $\varepsilon^3$.
These effects are generated by charm loop insertions and start 
at ${\cal O}(\als^3)$ in the first case [Fig. \ref{fig1loop}$b)$] 
and at ${\cal O}(\als^2)$ in the latter case [Fig. \ref{fig1loop}$a)$]. 
Therefore, considering a charm with a non-zero 
mass modifies the right-hand side of Eq. (\ref{spectrum}) 
(taken in the bottomonium case and with $n_l=4$ active and massless flavours) 
and of Eq. (\ref{pole})  (taken in the bottom case and with $n_l=4$ active 
and massless flavours) respectively by an amount
\begin{eqnarray}
(\delta E_{b\bar{b}})_{m_c} &=& \varepsilon^2 (\delta E_{b\bar{b}})^{(1)}_{m_c}
+ \varepsilon^3 (\delta E_{b\bar{b}})^{(2)}_{m_c}, \label{en}\\
(\delta m_b)_{m_c} &=& \varepsilon^2 (\delta m_b)^{(1)}_{m_c}
+ \varepsilon^3 (\delta m_b)^{(2)}_{m_c}. \label{m}
\end{eqnarray}
The term $(\delta E_{b\bar{b}})^{(1)}_{m_c}$ 
has been calculated for all quantum numbers in \cite{eiras};  
$(\delta m_b)^{(1)}_{m_c}$ has been calculated in \cite{gray}; 
$(\delta E_{b\bar{b}})^{(2)}_{m_c}$ has been calculated in \cite{hoangcm} 
for the $1S$ bottomonium level, but is unknown for higher quantum numbers;  
$(\delta m_b)^{(2)}_{m_c}$ has been calculated in \cite{hoangcm} in the limit
of $m_c \to 0$, keeping only the (leading) linear contribution 
in the charm mass (``linear approximation'').

In the following we will analyse these different contributions. Our main result will be the 
calculation of the dominant contribution to $(\delta E_{b\bar{b}})^{(2)}_{m_c}$ 
(at the level of 95\%) for all quantum numbers. 
Hence, we will be in a position to account for the finite charm-mass 
effects in the bottomonium spectrum up to order $\varepsilon^3$.

\begin{figure}[htb]
\makebox[0truecm]{\phantom b}
\put(10,0){\epsfxsize=7.2truecm\epsffile{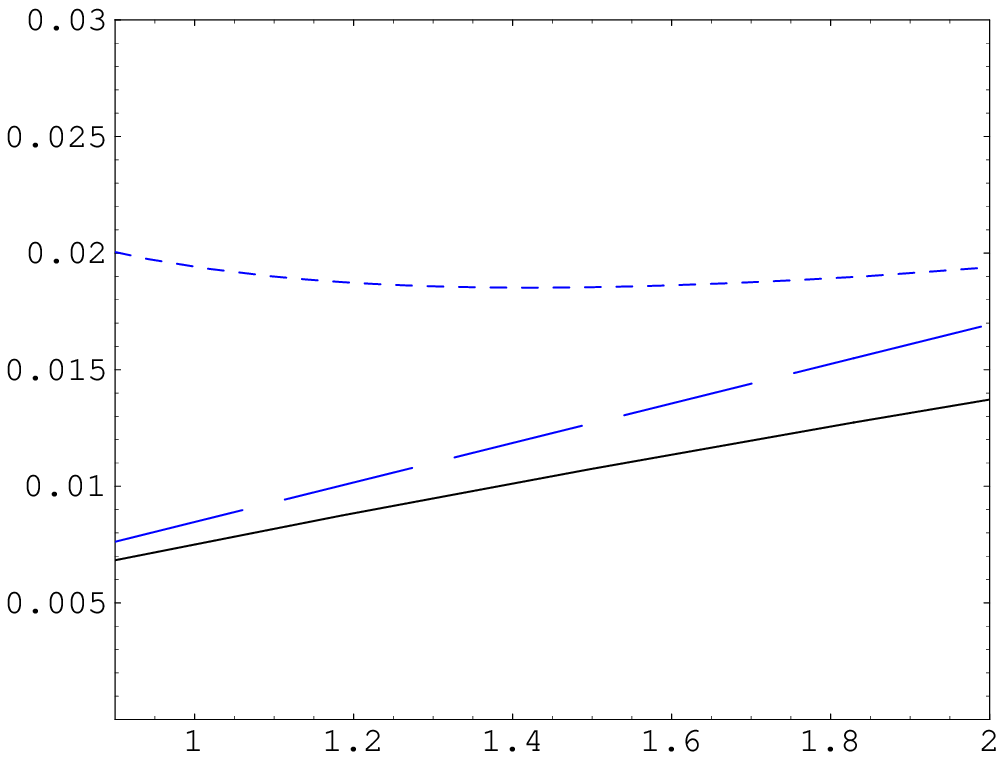}}
\put(250,0){\epsfxsize=7.2truecm\epsffile{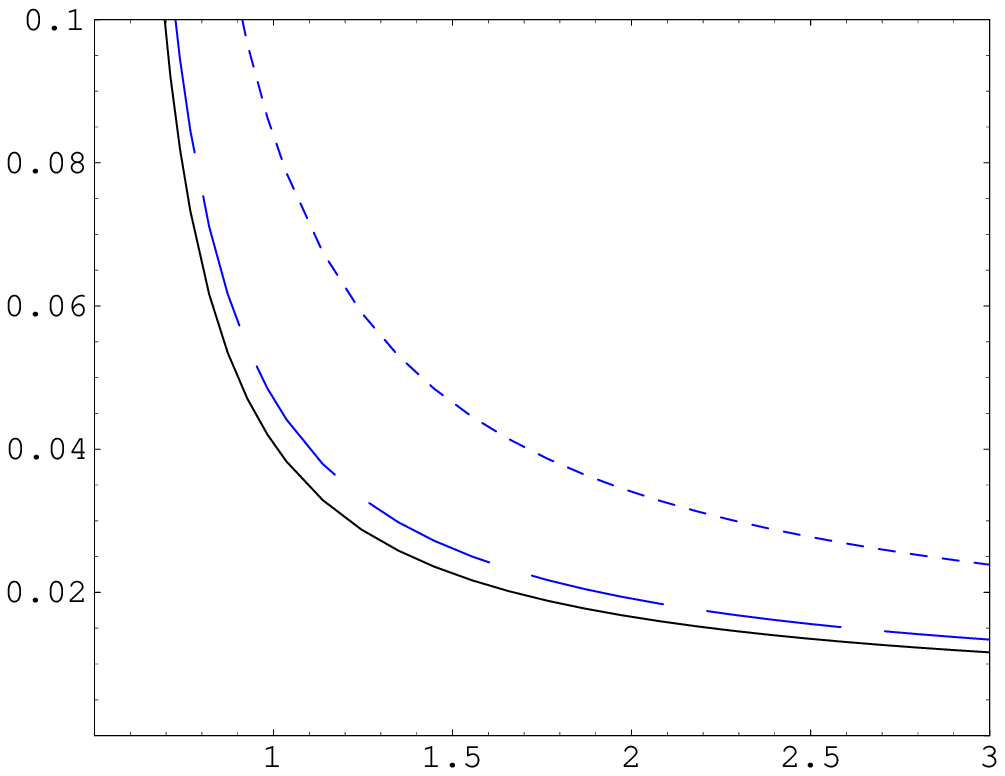}}
\put(200,-7){$\mbar_c$}
\put(440,-7){$\mu$}
\vspace{5mm}
\caption{\footnotesize\it
The first figure shows 
$(\delta m_b)^{(1)}_{m_c}$ (continuous line), 
$(\delta m_b)^{(1)}_{m_c\to 0}$ (dashed line), $(\delta m_b)^{(1)}_{m_c\to\infty}$ (dotted line), 
as a function of $\mbar_c$ for $\mbar_b = 4.201$ GeV. 
The second figure shows $(\delta m_b)^{(1)}_{m_c}$ (continuous line), 
$(\delta m_b)^{(1)}_{m_c\to0}$ (dashed line), $(\delta m_b)^{(1)}_{m_c\to\infty}$ (dotted line),
as a function of $\mu$, when $\als^{(4)}(\mbar_b)$ is replaced by $\als^{(4)}(\mu)$, for 
$\mbar_b = 4.201$ GeV and $\mbar_c = 1.237$ GeV. The units are GeV.}
\label{figdm1}
\end{figure}

\subsection{Order-$\varepsilon^2$ effects}
The ${\cal O}(\varepsilon^2)$ effects in Eq. (\ref{en}) and Eq. (\ref{m}) 
stem from the graphs $b)$ and $a)$, respectively, of Fig. \ref{fig1loop}. 
In \cite{gray} it was found that 
\begin{eqnarray}
(\delta m_b)^{(1)}_{m_c} &=& {\mbar_b\over 3}
\left({\als^{(4)}({\mbar_b})\over \pi}\right)^2 \left[
\ln^2(\xi) + {\pi^2\over 6} - \left(\ln(\xi) + {3\over2}\right)\xi^2 \right. \nn\\
& &\quad +(1+\xi)(1+\xi^3)\left({\rm Li}_2(-\xi)-{1\over 2}\ln^2(\xi)+\ln(\xi)\ln(1+\xi)
+{\pi^2\over 6}\right) \nn\\
& &\quad \left. 
+(1-\xi)(1-\xi^3)\left({\rm Li}_2(\xi)-{1\over 2}\ln^2(\xi)+\ln(\xi)\ln(1-\xi)
-{\pi^2\over 3}\right)\right],
\label{m1}
\end{eqnarray}
where $\xi =  {\mbar_c}/{\mbar_b}$. It is useful to define 
\begin{eqnarray}
(\delta m_b)^{(1)}_{m_c\to0} &=& 
{(\als^{(4)}({\mbar_b}))^2\over 6} {\mbar_c},  \label{m10} \\
(\delta m_b)^{(1)}_{m_c\to\infty} &=& 
{{\mbar_b}\over 3}\left({\als^{(4)}({\mbar_b})\over \pi}\right)^2
\left[ {151\over 72} + {\pi^2\over 6} + {13\over 6}\ln\left({{\mbar_c}\over {\mbar_b}}\right)
+ \ln^2\left({{\mbar_c}\over {\mbar_b}}\right)\right],
\label{m1oo}
\end{eqnarray}
where $(\delta m_b)^{(1)}_{m_c\to0}$ corresponds to the expansion of $(\delta m_b)^{(1)}_{m_c}$
for $m_c \to 0$ (up to ${\cal O}(m_c^2)$ corrections) and $(\delta
m_b)^{(1)}_{m_c\to\infty}$ corresponds to the expansion of  $(\delta
m_b)^{(1)}_{m_c}$ for  $m_c \to \infty$ (up to ${\cal O}(1/m_c^2)$ corrections).
In the first plot of Fig. \ref{figdm1} we show $(\delta m_b)^{(1)}_{m_c}$ as a
function of ${\mbar_c}$ for ${\mbar_b} = 4.201$ GeV (the value is taken from Eq. (\ref{bmass0})).
Here and in the following, if not specified differently, 
$\als^{(n_l)}$ is calculated from 
$\als^{(5)}(M_Z) = 0.1181$ (central values of Eqs. (\ref{l5})--(\ref{l3})).
In the region of interest, 1 GeV $\siml {\mbar_c} \siml$ 1.5 GeV, 
$(\delta m_b)^{(1)}_{m_c}$ turns out to be approximated 
by $(\delta m_b)^{(1)}_{m_c\to0}$ reasonably well, 
while $(\delta m_b)^{(1)}_{m_c\to\infty}$ is far off.  
More specifically at the value ${\mbar_c} =$ 1.237 GeV ($\xi = 0.294$), 
taken from Eq. (\ref{cmass0}), we have
\begin{equation}
(\delta m_b)^{(1)}_{m_c}          \simeq  9.1 \, {\rm MeV}, \qquad 
(\delta m_b)^{(1)}_{m_c\to0}      \simeq 10.5 \, {\rm MeV}, \qquad 
(\delta m_b)^{(1)}_{m_c\to\infty} \simeq 18.7 \, {\rm MeV} . 
\label{m1num}
\end{equation}
The error of the ``linear approximation'' is about 15\%, which agrees with the analysis done 
in \cite{hoangcm}. 

Since, in order to realize the renormalon cancellation, the coupling 
constant $\als^{(4)}(\mbar_b)$ has to be expanded around the scale $\mu$, in the second 
plot of Fig. \ref{figdm1} we show $(\delta m_b)^{(1)}_{m_c}$, $(\delta m_b)^{(1)}_{m_c\to0}$
and $(\delta m_b)^{(1)}_{m_c\to\infty}$ as functions of $\mu$ at  ${\mbar_c} =$ 1.237 GeV and 
${\mbar_b} =$ 4.201 GeV, when $\als^{(4)}(\mbar_b)$ is substituted by $\als^{(4)}(\mu)$ 
in Eqs. (\ref{m1})--(\ref{m1oo}). Also this plot confirms that $(\delta m_b)^{(1)}_{m_c\to0}$ 
approximates $(\delta m_b)^{(1)}_{m_c}$ reasonably well, while $(\delta m_b)^{(1)}_{m_c\to\infty}$ 
is far off.

\begin{figure}[thb]
\makebox[0truecm]{\phantom b}
\put(115,0){\epsfxsize=8.0truecm\epsffile{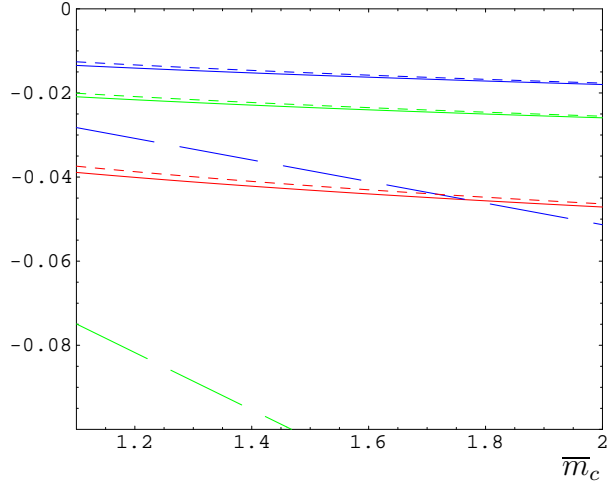}}
\put(325,-7){$\mbar_c$}
\vspace{5mm}
\caption{\footnotesize\it
$(\delta E)^{(1)}_{m_c}$ (continuous line), 
$(\delta E)^{(1)}_{m_c\to0}$ (dashed line), $(\delta E)^{(1)}_{m_c\to\infty}$ (dotted line), 
as a function of $\mbar_c$ for $\mbar_b = 4.201$ GeV. Going down, 
the first set of lines corresponds to the $1S$ state with $\mu = 2.446$ GeV, 
the second one to the $2S$ state with $\mu =  1.065$ GeV and the third one to 
the $3S$ state with $\mu = 0.724$ GeV. Lines, which are not displayed, fall outside the plot
range. The units are GeV.}
\label{figdE1}
\end{figure}

In \cite{eiras} it was found that 
\begin{eqnarray}
(\delta E_{b\bar{b}})^{(1)}_{m_c} &=& {{\mbar_b} 
(C_F\als^{(4)}(\mu))^2\over 4n^2}{\als^{(4)}(\mu)\over 3\pi}\left\{
- {3\pi\over 2}n\bar{\rho} + \bigg(n(2n+1)+(n+l)(n-l-1)\bigg)\bar{\rho}^2 \right.\nn\\
& & -\pi n \left({1\over 3}(n+1)(2n+1) + (n+l)(n-l-1)\right)\bar{\rho}^3 \nn\\
& & + 2 \ln\left({2\over \bar{\rho}}\right) - 2\bigg(\psi(n+l+1)-\psi(1)\bigg) \nn\\
& & - {2\over(2n-1)!}\sum_{k=0}^{n-l-1}
\left( 
\begin{array}{c} 
\displaystyle{n-l-1}\\ 
\displaystyle{k}\\ 
\end{array} 
\right)
\left( 
\begin{array}{c} 
\displaystyle{n+l}\\ 
\displaystyle{2l+1+k}\\ 
\end{array} 
\right)
\bar{\rho}^{2(n-l-1-k)} \nn\\
& & \left.\times {d^{2n-1}\over {d\bar{\rho}^{2n-1}}}\left[ 
\bar{\rho}^{2k+2l+1}{(2-\bar{\rho}^2-\bar{\rho}^4)\over\sqrt{\bar{\rho}^2-1}}
{\rm Atan}\left({\sqrt{\bar{\rho}-1}\over\sqrt{\bar{\rho}+1}}\right)\right]\right\},
\label{en1}
\end{eqnarray}
where $\bar{\rho} = 2 n {\mbar_c}/({\mbar_b} C_F \als^{(4)}(\mu))$. Again, it is useful to define 
\begin{eqnarray}
(\delta E_{b\bar{b}})^{(1)}_{m_c\to0} \!&=&\!\! 
- {C_F (\als^{(4)}(\mu))^2\over 4}{\mbar_c},\label{en10}\\
(\delta E_{b\bar{b}})^{(1)}_{m_c\to\infty} \! &=& \!\!
{{\mbar_b} (C_F\als^{(4)}(\mu))^2\over 4n^2}{\als^{(4)}(\mu)\over 3\pi}
\left\{2 \ln\left({2\over \bar{\rho}}\right) - {5\over 3} - 2\bigg(\psi(n+l+1)-\psi(1)\bigg) \right\},
\label{en1oo}
\end{eqnarray}
where $(\delta E_{b\bar{b}})^{(1)}_{m_c\to0}$ corresponds to the expansion of  
$(\delta E_{b\bar{b}})^{(1)}_{m_c}$ for  $m_c \to 0$ (up to ${\cal O}(m_c^2)$ corrections) and 
$(\delta E_{b\bar{b}})^{(1)}_{m_c\to\infty}$ corresponds 
to the expansion of  $(\delta E_{b\bar{b}})^{(1)}_{m_c}$ for  $m_c \to \infty$ 
(up to ${\cal O}(1/m_c^{2l+2})$ corrections).
In Fig. \ref{figdE1} we show $(\delta E_{b\bar{b}})^{(1)}_{m_c}$ as a function of ${\mbar_c}$ 
for ${\mbar_b} = 4.201$ GeV. We plot the functions (from up to down) 
corresponding to the $1S$ bottomonium level ($\mu = 2.446$ GeV), 
the $2S$ bottomonium level ($\mu = 1.065$ GeV), 
and the $3S$ bottomonium level ($\mu = 0.724$ GeV). The scales $\mu$ 
correspond to the extremum of $E_{b\bar{b}}$ (defined in Eq. (\ref{en2})) at ${\mbar_b} = 4.201$ GeV. 
It is noteworthy that $(\delta E_{b\bar{b}})^{(1)}_{m_c\to0}$ does not depend explicitly 
on the quarkonium state (the dependence in Fig. \ref{figdE1} is only due to the different scales).
In the region of interest 1 GeV $\siml {\mbar_c} \siml$ 1.5 GeV, 
$(\delta E_{b\bar{b}})^{(1)}_{m_c}$ turns out to be very  well approximated 
by $(\delta E_{b\bar{b}})^{(1)}_{m_c\to\infty}$ 
(for all the levels), while $(\delta E_{b\bar{b}})^{(1)}_{m_c\to0}$ is far off. 
More specifically in Table \ref{tabdE1} we list for all the levels at the above scales
the values of $(\delta E_{b\bar{b}})^{(1)}_{m_c}$, $(\delta E_{b\bar{b}})^{(1)}_{m_c\to0}$ 
and $(\delta E_{b\bar{b}})^{(1)}_{m_c\to\infty}$ with ${\mbar_c} =$ 1.237 GeV 
and ${\mbar_b} =$ 4.201 GeV. The error of the ``asymptotic approximation'' 
[$(\delta E_{b\bar{b}})^{(1)}_{m_c\to\infty}$] is extremely small. In the worst situation 
(the $1S$ level), it is about 5\%. In this case, however, the corrections are small 
(less than 1 MeV), so that they are beyond the accuracy 
of the present work. The ``linear approximation'' instead, at variance to what happens 
in the pole-mass expansion, is far off the exact values for all the states. 
This is in accordance with the general arguments of \cite{hoangcm}.
Let us notice, as a general remark,  that the ``asymptotic approximation'' for the 
energy levels, works better than the ``linear approximation'' for the pole-mass expansion.
We shall further comment on the above results in Sec. \ref{seccom}.
\begin{table}[ht]
\vspace{-2mm}
\makebox[2cm]{\phantom b}
\begin{tabular}{l||c|c|c|c|c|c}
\hline
State &$\mu$ &$\als^{(4)}(\mu)$ &$\bar{\rho}$ & $(\delta E_{b\bar{b}})^{(1)}_{m_c}$
& $(\delta E_{b\bar{b}})^{(1)}_{m_c\to0}$& $(\delta E_{b\bar{b}})^{(1)}_{m_c\to\infty}$ \\
\hline
$1 ^3S_1$ & 2.446 & 0.277& 1.59& $-$0.0143& $-$0.032& $-$0.0136\\
$1 ^3P_0$ & 1.140 & 0.428& 2.06& $-$0.0210& $-$0.076& $-$0.0210\\
$1 ^3P_1$ & 1.111 & 0.437& 2.02& $-$0.0221& $-$0.079& $-$0.0221\\
$1 ^3P_2$ & 1.086 & 0.445& 1.99& $-$0.0232& $-$0.082& $-$0.0232\\
$2 ^3S_1$ & 1.065 & 0.452& 1.96& $-$0.0219& $-$0.084& $-$0.0211\\
$2 ^3P_0$ & 0.726 & 0.695& 1.91& $-$0.0426& $-$0.199& $-$0.0424\\
$2 ^3P_1$ & 0.703 & 0.733& 1.81& $-$0.0490& $-$0.222& $-$0.0488\\
$2 ^3P_2$ & 0.678 & 0.782& 1.70& $-$0.0581& $-$0.252& $-$0.0579\\
$3 ^3S_1$ & 0.724 & 0.698& 1.90& $-$0.0405& $-$0.201& $-$0.0392\\
\hline
\end{tabular}
\vspace{5mm}
\caption{{\footnotesize\it
$(\delta E_{b\bar{b}})^{(1)}_{m_c}$, $(\delta E_{b\bar{b}})^{(1)}_{m_c\to0}$ and 
$(\delta E_{b\bar{b}})^{(1)}_{m_c\to\infty}$ for ${\mbar_b} =$ 4.201 GeV and 
${\mbar_c} =$ 1.237 GeV; $\als^{(4)}$ is calculated from $\Lambda_{\msbar}^{(4)} = 0.292$ GeV 
at four loops. All dimensionful quantities are expressed in GeV.}}
\vspace{2mm}
\label{tabdE1}
\end{table}

\begin{figure}[\protect{t}]
\makebox[0truecm]{\phantom b}
\put(10,0){\epsfxsize=7.2truecm\epsffile{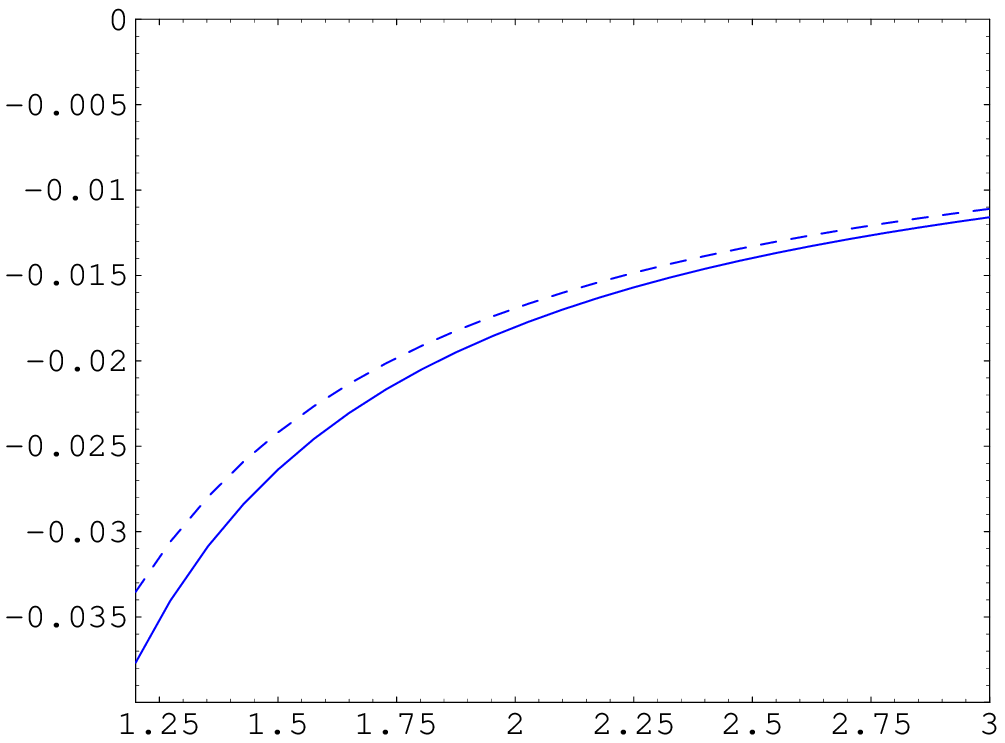}}
\put(250,0){\epsfxsize=7.2truecm\epsffile{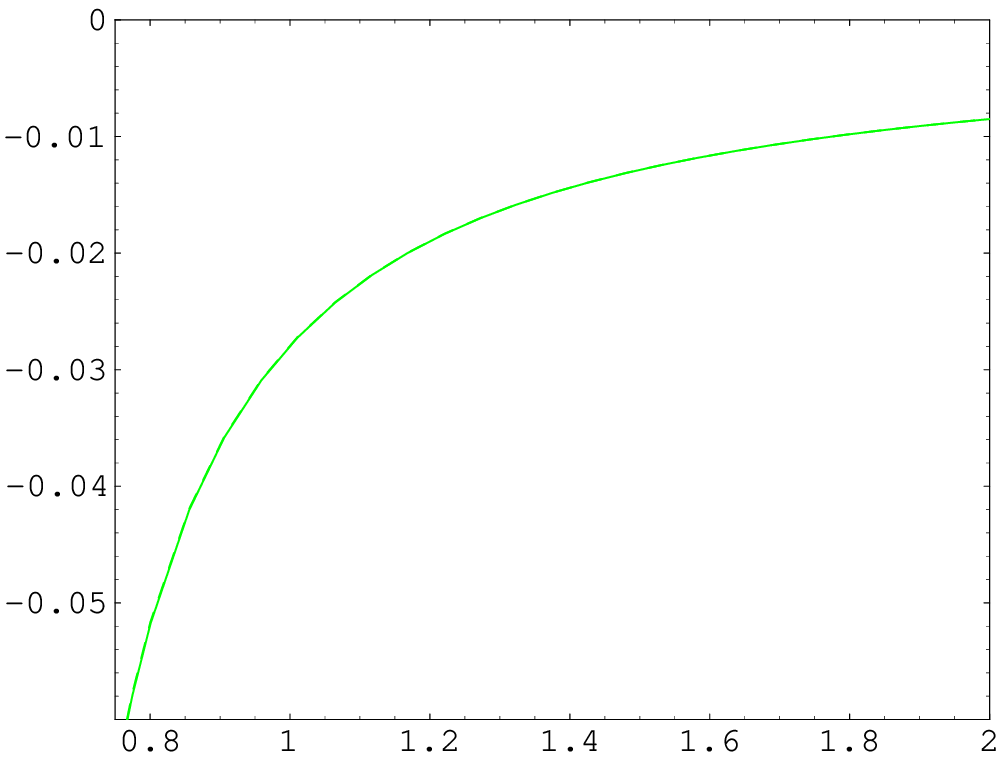}}
\put(10,-180){\epsfxsize=7.2truecm\epsffile{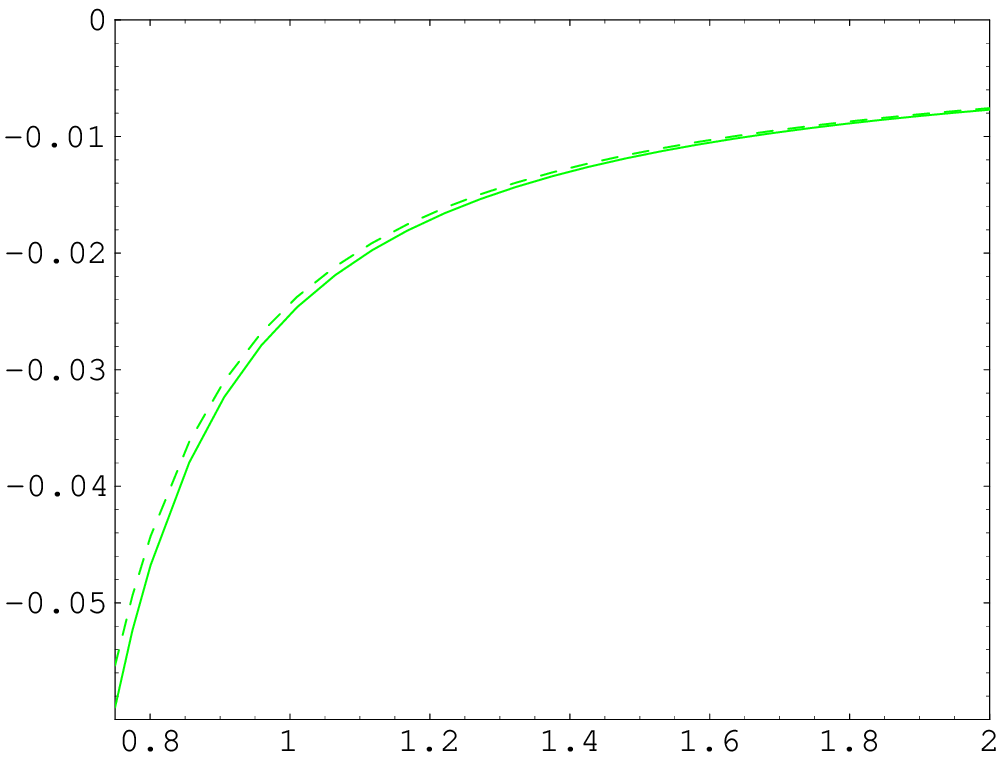}}
\put(250,-180){\epsfxsize=7.2truecm\epsffile{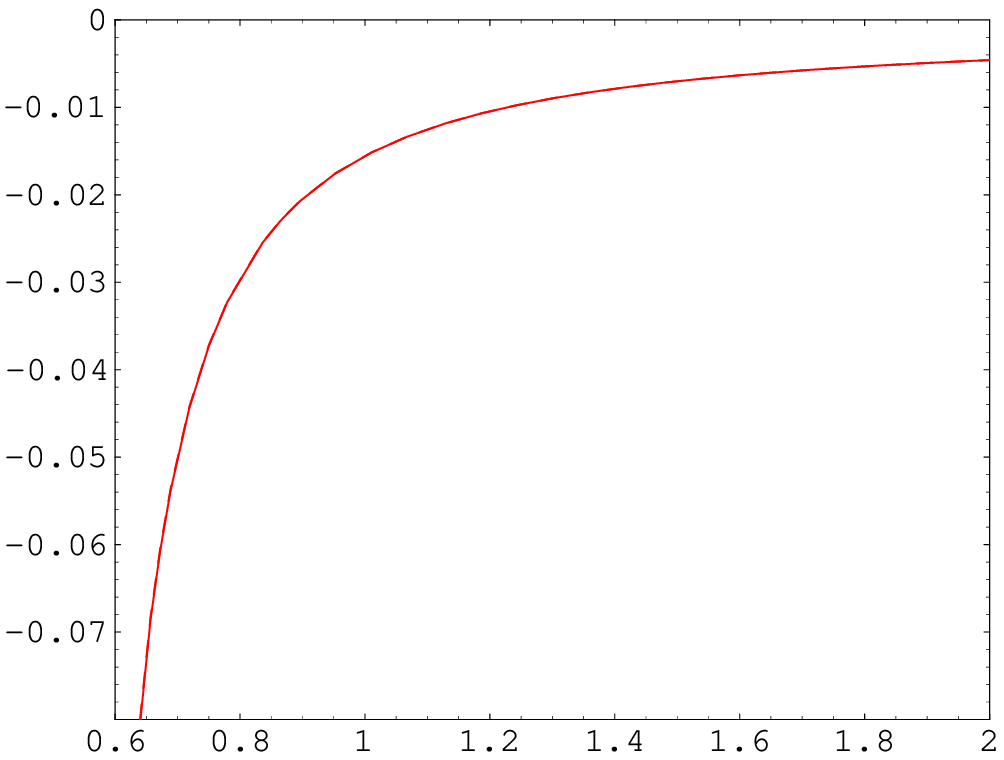}}
\put(10,-360){\epsfxsize=7.2truecm\epsffile{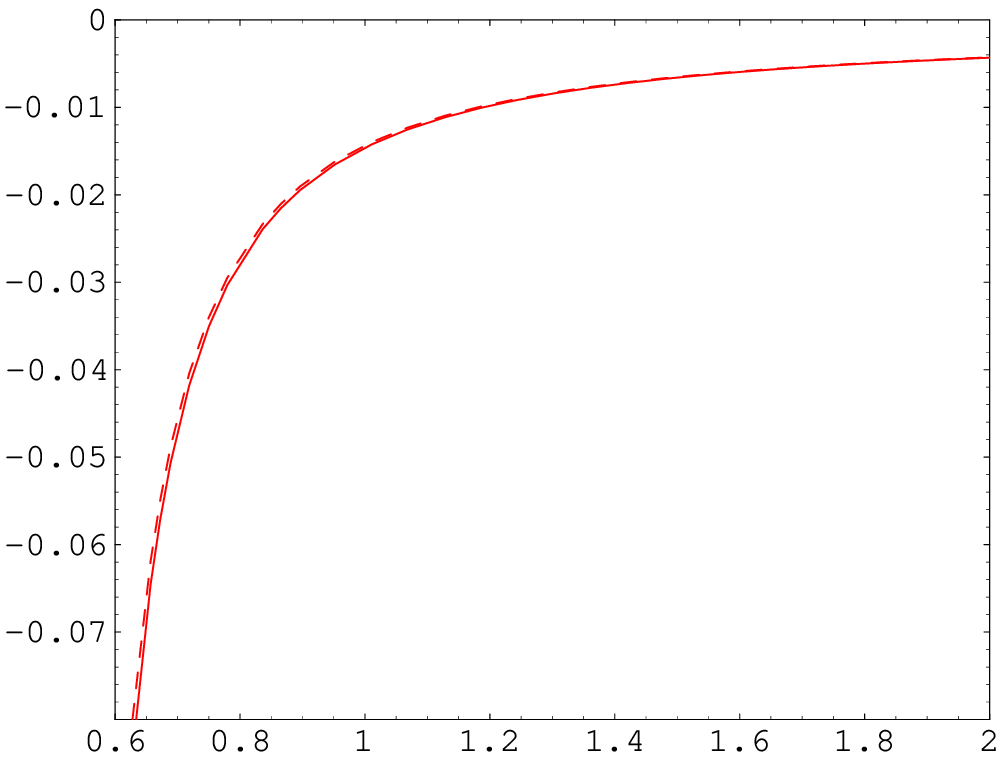}}
\put(250,-360){\epsfxsize=7.2truecm\epsffile{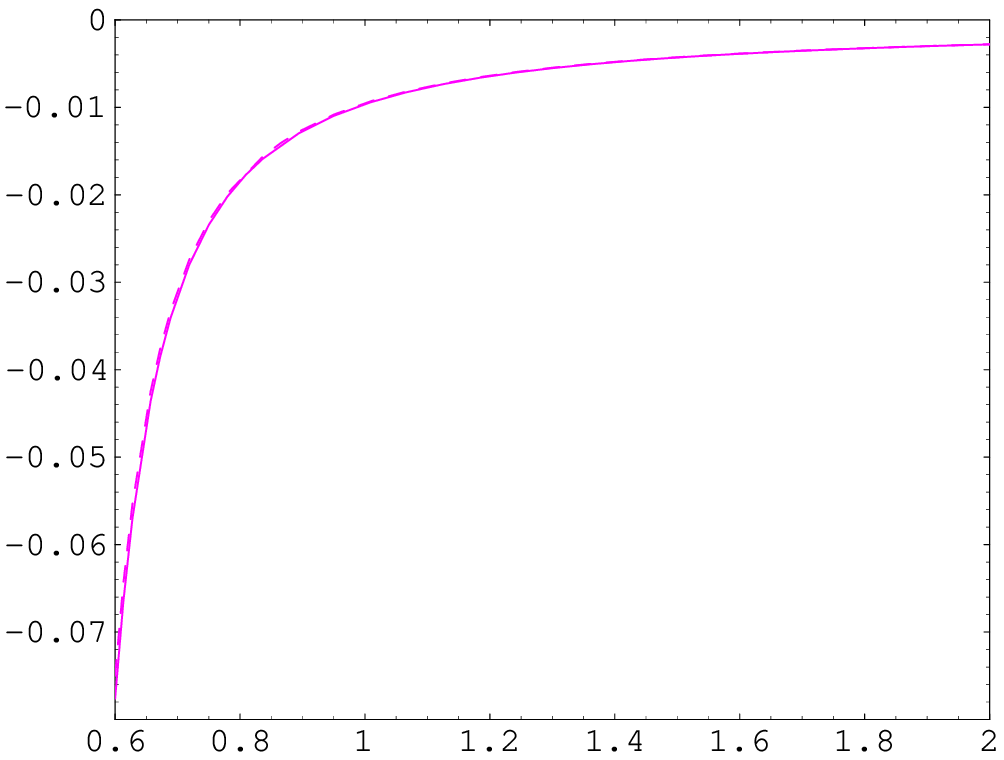}}
\put(50,135){$1S$}
\put(290,135){$1P$}
\put(50,-45){$2S$}
\put(290,-45){$2P$}
\put(50,-225){$3S$}
\put(290,-225){$4S$}
\put(200,-7){$\mu$}
\put(440,-7){$\mu$}
\put(200,-187){$\mu$}
\put(440,-187){$\mu$}
\put(200,-367){$\mu$}
\put(440,-367){$\mu$}
\vspace{5mm}
\caption{\footnotesize\it $(\delta E)^{(1)}_{m_c}$ (continuous line), 
and $(\delta E)^{(1)}_{m_c\to\infty}$ (dotted line), 
as a function of $\mu$ for ${\mbar_b}= 4.201$ GeV and ${\mbar_c} = 1.237$ GeV. 
The units are GeV.}
\label{figdE1mu}
\end{figure}

In Fig. \ref{figdE1mu} we show $(\delta E)^{(1)}_{m_c}$ (continuous line), 
and $(\delta E)^{(1)}_{m_c\to\infty}$ (dotted line), as a function of $\mu$ 
for ${\mbar_b}= 4.201$ GeV and ${\mbar_c} = 1.237$ GeV for all the bottomonium states.  
Also this figure confirms that $(\delta E)^{(1)}_{m_c\to\infty}$ approximates $(\delta E)^{(1)}_{m_c}$ 
very well for all the levels.

\subsection{Order-$\varepsilon^3$ effects}
\label{e3}
The $(\delta m_b)^{(2)}_{m_c}$ correction of Eq. (\ref{m}) is not known exactly. 
However, following the calculation of the two-loop mass
effects in the static potential in \cite{melles},
in \cite{hoangcm} $(\delta m_b)^{(2)}_{m_c\to0}$ has been calculated:
\begin{eqnarray}
(\delta m_b)^{(2)}_{m_c\to0} &=& 
{(\als^{(4)}({\mbar_b}))^3\over \pi} {\mbar_c}
\left\{ {2\over 9} + {\beta_0^{(4)}\over 12}\left(
2\ln\left({{\mbar_b}\over{\mbar_c}}\right) -4\ln 2 +{14\over 3}\right) \right.\nn\\
& & \qquad \left. -{1\over 9}\left({59\over 15}+2 \ln 2 \right) 
+ {19\over 9\pi}(f_1f_2 + b_1b_2)\right\},
\label{m20} 
\end{eqnarray}
where $f_2 = 0.470\pm0.005$, $b_2 = 1.120\pm0.010$,\footnote{
Somewhat larger errors are quoted in \cite{melles2}.
In our analysis only the central values are used.
} 
$f_1=(\ln A - \ln b_2)/(\ln f_2 -\ln b_2)$, 
$b_1=(\ln A - \ln f_2)/(\ln b_2 -\ln f_2)$ and $\ln A = 161/228 + 13\zeta_3/19 -\ln 2$. For the values
${\mbar_b} = 4.201$ GeV and ${\mbar_c} =$ 1.237 GeV we have
\begin{equation}
(\delta m_b)^{(2)}_{m_c\to0}      \simeq 17 \, {\rm MeV}. 
\label{m2num}
\end{equation}
From the analysis of the previous section we may expect that also this value approximates 
$(\delta m_b)^{(2)}_{m_c}$ with a relative uncertainty  smaller than 20\%. 
In \cite{hoangcm} it is argued that the sum $(\delta m_b)^{(1)}_{m_c\to0} + (\delta m_b)^{(2)}_{m_c\to0}$ 
may, indeed, approximate  $(\delta m_b)^{(1)}_{m_c} + (\delta m_b)^{(2)}_{m_c}$ with an uncertainty  
of 10\%, while the precision of the sum $(\delta m_b)^{(1)}_{m_c} + (\delta m_b)^{(2)}_{m_c\to0}$ 
is claimed to be worse. Finally, we notice that the series (\ref{m}) 
shows no signals of convergence (compare with the figures of Eq. (\ref{m1num})). 

The correction $(\delta E_{b\bar{b}})^{(2)}_{m_c}$ of Eq. (\ref{en}) is known only for the $1S$ quarkonium 
state \cite{hoangcm}.  Using the analytic expression given in  \cite{hoangcm} at the values of the masses 
${\mbar_b} = 4.201$ GeV, ${\mbar_c} = 1.237$ GeV and at the scale $\mu =
2.446$ GeV, we obtain\footnote{
We have used the $\msbar$ masses in the expression given in \cite{hoangcm} in terms of the 
pole masses. This introduces at order $\varepsilon^3$ also a correction equal to  
$4\als^{(4)}(\mu)/(3 \pi)\,(\delta E_{b\bar{b}})_{m_c}^{(1)}$ (see also 
Eq. (\ref{Emc2}) below). 
This correction amounts to $\simeq -1.7$ MeV for the full $(\delta E_{b\bar{b}})_{m_c}^{(1)}$ 
and to $\simeq -1.6$ MeV for the ``asymptotic approximation'' $(\delta E_{b\bar{b}})_{m_c\to\infty}^{(1)}$.}
\begin{equation}
(\delta E_{1S})^{(2)}_{m_c}          \simeq  -38.8  \, {\rm MeV}, \qquad 
(\delta E_{1S})^{(2)}_{m_c\to\infty} \simeq  -38.3  \, {\rm MeV} . 
\label{E2num}
\end{equation}
This confirms that the error of the ``asymptotic approximation'' is, indeed, extremely 
small (in the case of Eq. (\ref{E2num}) it is about 1\%). Considering that the $1S$ state is 
the state located furthest from the ``asymptotic'' limit, we may 
conclude that the uncertainty connected with the use of the ``asymptotic approximation'' in the 
energy expansion is for all levels below the accuracy of the present work and, hence,  negligible.
The explicit expression for $(\delta E_{b\bar{b}})^{(2)}_{m_c\to\infty}$ in terms of the pole mass 
is provided in Eq. (\ref{e2}) of \ref{ape}.

\subsection{Discussion}
\label{seccom}
In this section we interpret and discuss the above results. 
\vspace{2mm}\\
{\it a)} The terms $(\delta m_b)^{(1)}_{m_c}$ and $(\delta m_b)^{(2)}_{m_c}$ are governed by the 
parameter $\xi$, which is the ratio between the charm and the bottom $\msbar$ masses. This is natural,  
since these are the only scales involved in the pole mass expansion. Since $\xi \sim 0.3$, which is smaller 
than $1$, the small $\xi$, or $m_c$, approximation is expected to work. On closer inspection
of Eq. (\ref{m1}), we find 
\begin{equation}
{(\delta m_b)^{(1)}_{m_c} - (\delta m_b)^{(1)}_{m_c\to0} \over (\delta m_b)^{(1)}_{m_c}} = 
- {6\over\pi^2} \xi + {\cal O}(\xi^2) \simeq - 18\%,
\end{equation}
which practically accounts for the entire difference between the first two terms in (\ref{m1num}).
Therefore, we may say that, concerning the pole mass expansion, the charm 
mass can be considered small and we are close to the situation of four active and massless flavours.
\vspace{4mm}\\
{\it b)}  The terms $(\delta E_{b\bar{b}})^{(1)}_{m_c}$ 
and $(\delta E_{b\bar{b}})^{(2)}_{m_c}$ are governed by the 
parameter $\bar{\rho}$. This is twice the ratio of the charm $\msbar$ mass and the typical 
momentum of $b$ and $\bar{b}$, which is the relevant scale of the bound state. 
This number turns out to be typically larger than $1$ (see Table \ref{tabdE1}).  
Therefore, an expansion in small $\bar{\rho}$, or $m_c$, is out of the question and only an expansion 
for large $\bar{\rho}$, or $m_c$ may work. From Eq. (\ref{en1}) we have 
\begin{equation}
{(\delta E_{b\bar{b}})^{(1)}_{m_c} - (\delta E_{b\bar{b}})^{(1)}_{m_c\to\infty} \over (\delta E_{b\bar{b}})^{(1)}_{m_c}} \sim 
- {1\over \bar{\rho}^{2l+2}}{1\over 2\psi(n+l+1) -2\psi(1) +5/3}.
\end{equation}
It follows that, even if the expansion parameter $\bar{\rho}$ is not particularly 
large, the ``asymptotic approximation'' turns out to work very well 
(for all the states), since the leading discarded term is suppressed 
as $1/\bar{\rho}^{2l+2}$, i.e. by at least 
two powers in $1/\bar{\rho}$. The numerical factors contribute for extra suppressions.
This explains why the data of Table \ref{tabdE1} are reproduced so well by the 
``asymptotic approximation''. Notice that, {\it i)} states with higher $l$ are expected 
to be reproduced better by the asymptotic formula than states with lower $l$
(this can be checked explicitly in  Table \ref{tabdE1} by comparing the data 
for the $l=0$ states with those for the $l=1$ states, and in Fig. \ref{figdE1mu}); {\it ii)}
states with higher $n$ are expected to be reproduced better than states 
with lower $n$ (as long as we are in the perturbative regime), 
since the parameter $\bar{\rho}$ grows like $n/\als^{(4)}(\mu)$ (even 
if the growing of $\als^{(4)}(\mu)$ partially compensates the growing of $n$).  
In this sense the conclusions drawn in Sec. \ref{e3} on the relative size 
of $(\delta E_{b\bar{b}})^{(2)}_{m_c}$
with respect to $(\delta E_{b\bar{b}})^{(2)}_{m_c\to\infty}$ for the $1S$ state are expected to hold 
even better for the other (higher) states. 

We may conclude that, where the energy-level expansion (in terms of the pole mass) is concerned, 
the charm mass can be considered large with respect to the typical scale of
the bound state, so that the charm effectively decouples. 
Hence, charm-mass effects can be taken into account in the 
energy-level expansion of the bottomonium system (in terms of the  pole mass)
in a very effective way  by considering it in the situation with three 
active and massless flavours only\footnote{
It is worth while to stress that the situation here is the exact opposite with respect to that 
of the pole-mass expansion in terms of the $\msbar$ mass discussed in {\it a)}.}. We have 
\begin{equation}
(\delta E_{b\bar{b}})_{m_c} 
\hbox{ \raise-6pt\hbox{$\buildrel{\simeq} \over {\hbox{\tiny 95\%}}$}
}
(\delta E_{b\bar{b}})_{m_c\to\infty} 
= E_{{\rm bin},b\bar{b}}^{3}(\mu,\als^{(3)}(\mu),m_{b,{\rm pole}}) 
- E_{{\rm bin},b\bar{b}}^{4}(\mu,\als^{(4)}(\mu),m_{b,{\rm pole}}), 
\label{end}
\end{equation}
where $E_{{\rm bin},b\bar{b}}^{n_l}$ are the bottomonium binding energies in terms of the 
bottom pole mass (Eq. (\ref{spectrum2})) with $n_l$ active and massless 
flavours only. 
Equation (\ref{end}) is the key result of our analysis. It is expected to be valid 
at any order and it states that the charm decouples in the energy-level expansion 
(in terms of the pole mass) of the bottomonium system. Explicit expressions 
for Eq. (\ref{end}) up to order $\varepsilon^3$ are given in \ref{ape}.

\subsection{Cross-check}
\label{crosscheck}
We have made a non-trivial cross-check of the major result, Eq. (\ref{end}).
We compared the total static energy of a $b\bar{b}$ system,
$E_{\rm tot}(r) \equiv 2 m_{b, {\rm pole}} + V_{\rm QCD}(r)$, 
up to ${\cal O}(\als^3)$, in the case where the full charm-mass corrections 
are included in $V_{\rm QCD}(r)$, the static potential in perturbative QCD,  
and in that where $V_{\rm QCD}(r)$ is evaluated in the limit $m_c \to \infty$. 
To be more precise, we re-analysed the total 
energy $E_{\rm tot}(r)$, studied in two hypothetical cases in \cite{hep-ph/0104259}.
We have included the full charm-mass corrections to the static QCD potential 
using the formulas obtained in \cite{melles,hoangcm}.  Then we compared the total energy with 
the one calculated using $V_{\rm QCD}(r)$ in the limit $m_c \to \infty$, i.e. with $n_l=3$.  
[In both cases, $(\delta m_b)^{(i)}_{m_c}$ is approximated by $(\delta m_b)^{(i)}_{m_c\to 0}$.]
The total energies in the two cases agree with each 
other to a good accuracy: the difference is 3~MeV at $r=0.5~{\rm GeV}^{-1}$
and of order 0.1 MeV in the range $1~{\rm GeV}^{-1} < r < 3~{\rm GeV}^{-1}$.  
Since $E_{\rm tot}(r)$ determines the bulk of the bottomonium spectrum \cite{bsv1}, and since 
all the charm-mass effects are included in it, this confirms the validity of our approximation.
Note that the cross check was made {\it after} cancellation of the leading 
renormalon contributions (expressing $m_{b,{\rm pole}}$ in terms of
$\overline{m}_b$), which adds a non-trivial point to the analyses done in the previous sections.

\subsection{Level expansions in terms of $\msbar$ masses}
A way to implement the above results in the bottomonium level expansion in terms of the 
$\msbar$ masses is to modify Eq. (\ref{en2}) into
\bea
& & \hspace{-5mm} E_{b\bar{b}}
= 
\left[
2 \overline{m}_b \left\{ 1 + {4\over 3} \varepsilon {\als^{(4)}(\overline{m}_b)\over \pi} 
+ \varepsilon^2  \left({\als^{(4)}(\overline{m}_b)\over \pi}\right)^2 d^{(4)}_1 
+ \varepsilon^3 \left({\als^{(4)}(\overline{m}_b)\over \pi}\right)^3 d^{(4)}_2 \right\} \right.
\nn\\
& & \quad \>
+ E^{3}_{{\rm bin},b\bar{b}} \left(\mu,\als^{(3)}(\mu),
\overline{m}_b \left\{ 1 + {4\over 3} \varepsilon {\als^{(4)}(\overline{m}_b)\over \pi} 
+ \varepsilon^2  \left({\als^{(4)}(\overline{m}_b)\over \pi}\right)^2 d^{(4)}_1 \right\} 
\right) \nn\\
& & \quad\> \left.
+ \varepsilon^2 2 (\delta m_b)^{(1)}_{m_c} 
+ \varepsilon^3 \left\{  2 (\delta m_b)^{(2)}_{m_c} 
- {1\over 4}\left({C_F \als^{(4)}(\mu)\over n}\right)^2 (\delta m_b)^{(1)}_{m_c} \right\} 
\right]_{\begin{array}{l}
\hbox{ \footnotesize $\!\!\als(\mbar_b)=$Eq. (\ref{alphams})} \\  
\hbox{ \footnotesize $\!\!\als^{(4)}=$Eq. (\ref{alphams2})} 
\end{array} }
\label{Emc} \\
& & \nn \\
& & \>\equiv
2 \overline{m}_b
+ E_{b\bar{b}}^{(1)}(\mu, \als^{(3)}(\mu), \mbar_c, \overline{m}_b) \varepsilon 
+ E_{b\bar{b}}^{(2)}(\mu, \als^{(3)}(\mu), \mbar_c, \overline{m}_b) \varepsilon^2 \nn\\
& & \quad \>
+ E_{b\bar{b}}^{(3)}(\mu, \als^{(3)}(\mu), \mbar_c, \overline{m}_b) \varepsilon^3 + \dots,
\label{Emcbis}
\eea
where $\als^{(4)}(\mbar_b)$ in Eq. (\ref{Emc})
is understood as expanded everywhere around the scale $\mu$ according to
Eq. (\ref{alphams}) and around the coupling with $3$ active massless flavours
according to the following equation \cite{running}:
\bea 
\als^{(4)}(\mu) &=& \als^{(3)}(\mu)\left\{ 
1 + \varepsilon \, \frac{\als^{(3)}(\mu)}{3 \pi}  \, 
\ln\left(\frac{\mu}{\overline{m}_c}\right) \right. \nn\\
& &\qquad\qquad\quad   \left.
+ \varepsilon^2 \left( \frac{\als^{(3)}(\mu)}{\pi} \right)^2 \,
\Biggl[ {1\over 9} \, \ln^2 \left(\frac{\mu}{\overline{m}_c}\right)
+ {19\over 12}\ln\left(\frac{\mu}{\overline{m}_c}\right) - {11\over 72}\Biggr] \right\}.
\label{alphams2}
\eea
Both expansions are needed in order to have the same coupling in the pole mass
and in the binding energy expansion so that the 
${\cal O}(\Lambda_{\rm QCD})$ renormalon cancellation is made explicit. 
Finite charm-mass corrections affect the pole-mass expansion and are explicitly written 
in the last line of Eq. (\ref{Emc}).

Another possibility is to express also the binding energy at four massless flavours and to 
calculate the corrections to it. In this case Eq. (\ref{en2}) is modified into 
\bea
& & \hspace{-5mm} E_{b\bar{b}} =  \left[
2 \overline{m}_b \left\{ 1 + {4\over 3} \varepsilon {\als^{(4)}(\overline{m}_b)\over \pi} 
+ \varepsilon^2  \left({\als^{(4)}(\overline{m}_b)\over \pi}\right)^2 d^{(4)}_1 
+ \varepsilon^3 \left({\als^{(4)}(\overline{m}_b)\over \pi}\right)^3 d^{(4)}_2 \right\} \right. \nn\\
& & \quad \>
+ E^{4}_{{\rm bin},b\bar{b}} \left(\mu,\als^{(4)}(\mu),
\overline{m}_b \left\{ 1 + {4\over 3} \varepsilon {\als^{(4)}(\overline{m}_b)\over \pi} 
+ \varepsilon^2  \left({\als^{(4)}(\overline{m}_b)\over \pi}\right)^2 d^{(4)}_1 \right\} \right) \nn\\
& & \quad \>
+ \varepsilon^2 \bigg\{  2 (\delta m_b)^{(1)}_{m_c} 
+ (\delta E_{b\bar{b}})^{(1)}_{m_c}({\mbar_b},\bar{\rho})  \bigg\} \nn\\
& & \quad \>
+ \varepsilon^3 \left\{  2 (\delta m_b)^{(2)}_{m_c} 
- {1\over 4} \left({C_F \als^{(4)}(\mu)\over n}\right)^2 (\delta m_b)^{(1)}_{m_c} \right.
\nn\\
& & \qquad\qquad
\left.\left.
+ (\delta E_{b\bar{b}})^{(2)}_{m_c}({\mbar_b},\bar{\rho}) 
+ {4\over 3}{\als^{(4)}(\mu)\over \pi} (\delta E_{b\bar{b}})^{(1)}_{m_c}({\mbar_b},\bar{\rho}) \right\}
\right]_{\als(\mbar_b) = \hbox{\footnotesize Eq. (\ref{alphams})}}
\label{Emc2} \\
& & \nn \\
& & \>
\equiv 
2\overline{m}_b 
+ E_{b\bar{b}}^{(1)}(\mu, \als^{(4)}(\mu), \mbar_c,  \overline{m}_b) \varepsilon 
+ E_{b\bar{b}}^{(2)}(\mu, \als^{(4)}(\mu),  \mbar_c, \overline{m}_b) \varepsilon^2 \nn\\
& & \quad \>
+ E_{b\bar{b}}^{(3)}(\mu, \als^{(4)}(\mu), \mbar_c, \overline{m}_b) \varepsilon^3 + \dots,  
\label{Emc2bis}
\eea
where we have made use of the fact that $\rho = \bar{\rho}(1+{\cal O}(\als^2))$,
and where $\als^{(4)}(\mbar_b)$ in Eq. (\ref{Emc2}) is understood as expanded everywhere around 
the scale $\mu$ according to Eq. (\ref{alphams}). The first two lines of Eq. (\ref{Emc2}) exactly 
correspond to Eq. (\ref{en2}), when calculated in the bottomonium case with four massless flavours.

Let us summarize our present knowledge of the different pieces of Eqs. (\ref{Emc}) and (\ref{Emc2}).

{\it A)}
$E^{n_l}_{{\rm bin},b\bar{b}}$ are the bottomonium binding energies in terms of the $\msbar$ 
masses with $n_l$ active and massless flavours only (i.e. $m_c=0$). They can be read from Eq. (\ref{spectrum2}).
A study of these contributions alone can be found in Ref. \cite{bsv1}.

{\it B)}
$(\delta m_b)^{(1)}_{m_c}$ is given in Eq. (\ref{m1}). If we expand $\als^{(4)}({\mbar_b})$ 
around the scale $\mu$, we then have to apply the replacement of Eq. (\ref{alphams}).
If we expand $\als^{(4)}(\mu)$ around  $\als^{(3)}(\mu)$, we then have to
apply also the replacement of Eq. (\ref{alphams2}).
The quantity $(\delta m_b)^{(2)}_{m_c}$ is known only in the small $m_c$ 
limit and is given in Eq. (\ref{m20}). 
The use of $(\delta m_b)^{(2)}_{m_c\to0}$ instead of $(\delta m_b)^{(2)}_{m_c}$ should account 
for 80\% of the effect. In \cite{hoangcm} it is claimed that the simultaneous use of 
$(\delta m_b)^{(1)}_{m_c\to0}$ and $(\delta m_b)^{(2)}_{m_c\to0}$ should account for 90\% 
of the actual effect. 

{\it C)}
$(\delta E_{b\bar{b}})^{(1)}_{m_c}({\mbar_b},\bar{\rho})$ is given in Eq. (\ref{en1}). 
The large $m_c$ limit of this expression is given in Eq. (\ref{en1oo}) and practically 
accounts for the full effect. The term $(\delta
E_{b\bar{b}})^{(2)}_{m_c}({\mbar_b},\bar{\rho})$ has been 
calculated here in the large $m_c$ limit and is given in Eq. (\ref{e2}). 
For the reasons discussed in Secs. \ref{seccom} and \ref{crosscheck}, 
we believe that Eq. (\ref{e2}) accounts for practically
the full effect. 

Since Eq. (\ref{Emc}) accounts explicitly for the decoupling of the charm quark at all orders in the 
binding-energy expansion (in terms of the pole mass), we expect results in
this case to be more stable and reliable than by using Eq. (\ref{Emc2}). 
This will be confirmed by our numerical analysis.
Therefore, we shall use Eq. (\ref{Emc}) as our reference calculation (Sec. \ref{sanab}). 
Equation (\ref{Emc2}) will be used in our analysis of theoretical uncertainties (Sec. \ref{ser}).
Moreover, in our reference calculation of Sec. \ref{sanab} we will approximate 
both $(\delta m_b)^{(1)}_{m_c}$ and $(\delta m_b)^{(2)}_{m_c}$ by means of the 
``linear approximations'' $(\delta m_b)^{(1)}_{m_c\to0}$ and 
$(\delta m_b)^{(2)}_{m_c\to0}$. This proves to be slightly more stable than the use 
of the full correction  $(\delta m_b)^{(1)}_{m_c}$. We take into account 
the full correction in the analysis of the uncertainties in Sec. \ref{ser}.

We conclude this section by giving the expression, 
which substitutes Eq. (\ref{en2}) in the $B_c$ case, 
when finite charm-mass effects are taken into account:
\bea
& & \hspace{-5mm} E_{B_c} =  \left[ \mbar_c 
\left\{ 1 + {4\over 3} \varepsilon {\als^{(3)}(\overline{m}_c)\over \pi} 
+ \varepsilon^2  \left({\als^{(3)}(\overline{m}_c)\over \pi}\right)^2 d^{(3)}_1 
+ \varepsilon^3 \left({\als^{(3)}(\overline{m}_c)\over \pi}\right)^3 d^{(3)}_2 \right\} \right.\nn\\
& & \hspace{-2mm}
+ \overline{m}_b \left\{ 1 + {4\over 3} \varepsilon {\als^{(4)}(\overline{m}_b)\over \pi} 
+ \varepsilon^2  \left({\als^{(4)}(\overline{m}_b)\over \pi}\right)^2 d^{(4)}_1 
+ \varepsilon^3 \left({\als^{(4)}(\overline{m}_b)\over \pi}\right)^3 d^{(4)}_2 \right\} \nn\\
& & \hspace{-2mm}
+ E^{3}_{{\rm bin},B_c} \left(\mu,\als^{(3)}(\mu),
\mbar_c
\left\{ 1 + {4\over 3} \varepsilon {\als^{(3)}(\overline{m}_c)\over \pi} 
+ \varepsilon^2  \left({\als^{(3)}(\overline{m}_c)\over \pi}\right)^2 d^{(3)}_1 \right\},\right. \nn\\
& & \qquad\qquad\qquad\qquad 
\left.
\mbar_b 
\left\{ 1 + {4\over 3} \varepsilon {\als^{(4)}(\overline{m}_b)\over \pi} 
+ \varepsilon^2  \left({\als^{(4)}(\overline{m}_b)\over \pi}\right)^2 d^{(4)}_1 \right\} \right) \nn\\
& & \left.\hspace{-2mm}
+ \varepsilon^2 (\delta m_b)^{(1)}_{m_c} 
+ \varepsilon^3 \left\{  (\delta m_b)^{(2)}_{m_c} 
- {1\over 2}\left({C_F \als^{(4)}(\mu)\over n}\right)^2 \!\!
\left({\mbar_c\over \mbar_c+\mbar_b}\right)^2 \!\! (\delta m_b)^{(1)}_{m_c} \right\} 
\right]_{\begin{array}{l}
\hbox{ \footnotesize $\!\!\als(\mbar_{c,b})=$Eq. (\ref{alphams})} \\  
\hbox{ \footnotesize $\!\!\als^{(4)}=$Eq. (\ref{alphams2})} 
\end{array} }
\nn \\
& & \label{EBc} \\
& & \hspace{-2mm} \equiv 
\overline{m}_c + \overline{m}_b 
+ E_{B_c}^{(1)}(\mu, \als^{(3)}(\mu), \mbar_c, \overline{m}_b) \varepsilon 
+ E_{B_c}^{(2)}(\mu, \als^{(3)}(\mu), \mbar_c, \overline{m}_b) \varepsilon^2 \nn\\
& & \hspace{-2mm}
+ E_{B_c}^{(3)}(\mu, \als^{(3)}(\mu), \mbar_c, \overline{m}_b) \varepsilon^3 + \dots, 
\label{EBcbis}
\eea
where $\als^{(3)}(\mbar_c)$ and $\als^{(4)}(\mbar_b)$ in Eq. (\ref{EBc}) are
understood as expanded everywhere around the scale $\mu$ according to
Eq. (\ref{alphams}) and $\als^{(4)}(\mu)$ around  $\als^{(3)}(\mu)$
according to Eq. (\ref{alphams2}).

\section{Numerical Analyses}
\label{san}

In this section we examine numerically the series expansions 
of the bottomonium spectrum, Eqs. (\ref{Emc}) and (\ref{Emc2}), 
and of the $B_c$ mass, Eq. (\ref{EBc}). 

\subsection{Scale-fixing procedures}
\label{secfix}
The quarkonium mass $E_X$ depends on the scale $\mu$, because of 
our incomplete knowledge of the perturbative series. 
We will fix the scale $\mu$ for each state $X$ in two different ways.
\begin{itemize}
\item[{\it A})]
We fix the scale $\mu=\mu_X^A$ by demanding stability of $E_X$ against variation
of the scale:
\bea
\left.\frac{d}{d\mu} E_{X}(\mu, \als^{(n_l)}(\mu),\overline{m}_i)\right|_{\mu = \mu_X^A} = 0 .
\label{scalefixA} 
\eea
\item[{\it B})]
We fix the scale $\mu=\mu_X^B$ on the minimum of 
$|E_{X}^{(3)}|$ :
\bea
\left.\frac{d}{d\mu} 
[E_{X}^{(3)}(\mu, \als^{(n_l)}(\mu),\overline{m}_i)]^2 \right|_{\mu = \mu_X^B} = 0 . 
\label{scalefixB} 
\eea
\end{itemize}
When we do this, we expect that the convergence properties of the series become optimal, and
that the scale becomes close to the inverse of the physical size, $a_X$, of the
bound state $X$ (defined as in Ref. \cite{bsv1}).
If the scales fixed by Eq. (\ref{scalefixA}) and Eq. (\ref{scalefixB}) 
evidently do not fulfil these expectations, the theoretical predictions obtained in this way 
will be considered as {\it unreliable}. This typically happens when 
the coupling constant becomes larger than 1. In the next section we will check 
if  $|E_{X}^{(1)}| > |E_{X}^{(2)}| > |E_{X}^{(3)}|$ at the chosen scales.  
Notice that the criterion $B)$ of Eq. (\ref{scalefixB}) practically always 
fulfils this request since ${\rm min} \{ |E_{X}^{(3)}| \} =0$ in all the considered cases. 
This means that the results obtained using criterion $B)$ have to be taken 
with some caution. For this reason, in the following we will use them, rather
than as an independent way to calculate the heavy-quarkonium spectra, as an independent 
check of the results obtained using criterion $A)$ as well as a means to determine 
the uncertainties related to the fixing of the scale $\mu_X$.
Comparisons of the chosen scales and the bound-state sizes have been given
in Sec. 5 of \cite{bsv1} and we do not repeat them here.

\begin{table}[t]
\begin{center}
\begin{tabular}{c||r|r|r|r|r|r|c|c}
\hline
State $X$ &$E_X^{\rm exp}~~$ &$E_X~~~$ & $E_X^{\rm exp}-E_X$ 
&$E_X^{(1)}$ &$E_X^{(2)}$ &$E_X^{(3)}$ &~~$\mu_X^A$ &$\als^{(3)}(\mu_X^A)$ \\ 
\hline
$\Upsilon(1^3S_1)$  & 9.460 & 9.460~~~~ & 0~~~~~ & 0.866 & 0.208  & 0.006 & 2.14 & 0.286 \\
$\Upsilon(1 ^3P_0)$ & 9.860 & $9.995^{+75}_{-62}$ & $-$0.135$^{-75}_{+62}$ & 1.534  & 0.101  &$-$0.021 & 1.08 & 0.459 \\
$\Upsilon(1 ^3P_1)$ & 9.893 &$10.004^{+78}_{-63}$ & $-$0.111$^{-78}_{+63}$ & 1.564  & 0.081  &$-$0.022 & 1.05 & 0.468 \\
$\Upsilon(1 ^3P_2)$ & 9.913 &$10.012^{+81}_{-65}$ & $-$0.099$^{-81}_{+65}$  & 1.591  & 0.063  &$-$0.022 & 1.034 & 0.477 \\
$\Upsilon(2 ^3S_1)$ &10.023 &$10.084^{+93}_{-75}$ & $-$0.061$^{-93}_{+75}$ & 1.618  & 0.096  &$-$0.010 & 1.02 & 0.486 \\
$\Upsilon(2 ^3P_0)$ &10.232 &$10.548^{+196}_{-151}$ & $-$0.316$^{-196}_{+151}$  & 2.421  &$-$0.356  & 0.102 & 0.778 & 0.710 \\
$\Upsilon(2 ^3P_1)$ &10.255 &$10.564^{+200}_{-153}$ & $-$0.309$^{-200}_{+153}$ & 2.472  &$-$0.404  & 0.116 & 0.770 & 0.726 \\
$\Upsilon(2 ^3P_2)$ &10.268 &$10.578^{+203}_{-155}$ & $-$0.310$^{-203}_{+155}$ & 2.518  &$-$0.449  & 0.129 & 0.762 & 0.740 \\
$\Upsilon(3 ^3S_1)$ &10.355 &$10.645^{+218}_{-168}$ & $-$0.290$^{-218}_{+168}$  & 2.472  &$-$0.348  & 0.140 & 0.770 & 0.726 \\
$\Upsilon(4 ^3S_1)$ &10.580 &$*$                    & $*$ & $*$  &$*$ & $*$ & $*$ & $*$ \\
$B_c(1^1S_0)$ & $6.4 \pm 0.4$ & $6.307^{+4}_{-2}$ & 0.1$\pm 0.4$ & 0.675 & 0.188 & 0.017 & 1.62 & 0.334 \\
\hline
\end{tabular}
\end{center}
\vspace{1mm}
\caption{\footnotesize\it
Theoretical predictions for the bottomonium and $B_c$ masses 
from Eqs. (\ref{Emc}) and 
(\ref{EBc}), respectively ($(\delta m_b)^{(1,2)}_{m_c} \simeq (\delta m_b)^{(1,2)}_{m_c\to0}$), 
when the scales $\mu_X^A$ are fixed through Eq. (\ref{scalefixA}).
The quantities $E_X^{(j)}$ are the order $\varepsilon^j$ contributions to the spectrum and 
are defined by Eqs. (\ref{Emcbis}) and (\ref{EBcbis}).
The $c$-  and $b$-quark $\msbar$ masses are fixed on the experimental 
values of the $J/\psi$ and $\Upsilon(1S)$ 
masses, respectively. The coupling constant $\als^{(3)}$ has been defined according to
Sec. \ref{parameter}. The uncertainties in the third and fourth columns refer 
to the uncertainties in $\als^{(5)}(M_Z)$ only. All the other data refer to
$\als^{(5)}(M_Z)=0.1181$ and to the $\msbar$ quark 
masses fixed on the central values of Eqs. (\ref{cmass}) and (\ref{bmass}). 
All dimensionful numbers are in GeV.}
\label{table:spectraA}
\end{table}

\subsection{Numerical results}
\label{sanab}
Here, we present results from both scale-fixing procedures on Eq. (\ref{Emc})
for the bottomonium spectrum and on Eq. (\ref{EBc}) for the $B_c$ mass. 
We calculate both $(\delta m_b)^{(1)}_{m_c}$ and $(\delta m_b)^{(2)}_{m_c}$ by means of the 
``linear approximations'' $(\delta m_b)^{(1)}_{m_c\to0}$ and $(\delta m_b)^{(2)}_{m_c\to0}$. 
The coupling constant $\als$ is defined as in Sec. \ref{parameter}.
The uncertainties on $\als^{(5)}(M_Z)$ [or on $\Lambda_\msbar$, see Eqs. (\ref{l5})--(\ref{l3})]
are the only ones that we take into account here.
Other sources of uncertainties and in particular those related to the finite charm-mass 
effects will be discussed in Sec. \ref{ser}.

\noindent
{\it Scale-fixing procedure A)}\vspace{2mm}\\
Since we expect the ground states of the $b\bar{b}$ and $c\bar{c}$
systems to be the states least affected by non-perturbative corrections,
we fix $\overline{m}_b$ and $\overline{m}_c$ through the two conditions
\begin{eqnarray}
E_{\Upsilon(1S)} (\mu_X^A, \als^{(3)}(\mu_X^A), \mbar_c, \mbar_b)
&=& E_{\Upsilon(1S)}^{\rm exp} = 9.460 \, \hbox{GeV} 
\label{con1},\\
E_{J/\psi} (\mu_X^A, \als^{(3)}(\mu_X^A),\mbar_c) &=& E_{J/\psi}^{\rm exp} = 3.097 \, \hbox{GeV} , 
\label{con2}
\end{eqnarray}
where the experimental values of the vector ground states have been taken from \cite{pdg}.  
We assume, for the moment, that this identification is not affected by non-perturbative corrections. 
From Eqs. (\ref{con1}), (\ref{con2}) and (\ref{scalefixA}) we determine $\mu_X^A$ 
(see Table \ref{table:spectraA}), and the $b$ and $c$ $\overline{\rm MS}$ masses:
\bea
\overline{m}_b &=&4190^{-20}_{+19} ~ {\rm MeV} ,
\label{bmass}
\\
\overline{m}_c &=& 1237^{-16}_{+16} ~ {\rm MeV} .
\label{cmass}
\eea
These values are in good agreement with the estimates based on $\Upsilon$ 
\cite{pp,my,upsilonmass,hoangcm} and charmonium \cite{jamin} sum rules respectively.  
The charm mass given in Eq. (\ref{cmass}) corresponds 
to the one quoted in Table 2, column (i) of Ref. \cite{bsv1} 
(reported also in Eq. (\ref{cmass0}) of the present paper), 
since it is not affected in our analysis by massive quark-loop effects. 
The bottom mass given in Eq. (\ref{bmass}) is new. 
The errors refer to the uncertainties in $\als^{(5)}(M_Z)$ only. 

\begin{figure}[\protect{h}]
\makebox[0truecm]{\phantom b}
\put(10,0){\epsfxsize=7.2truecm\epsffile{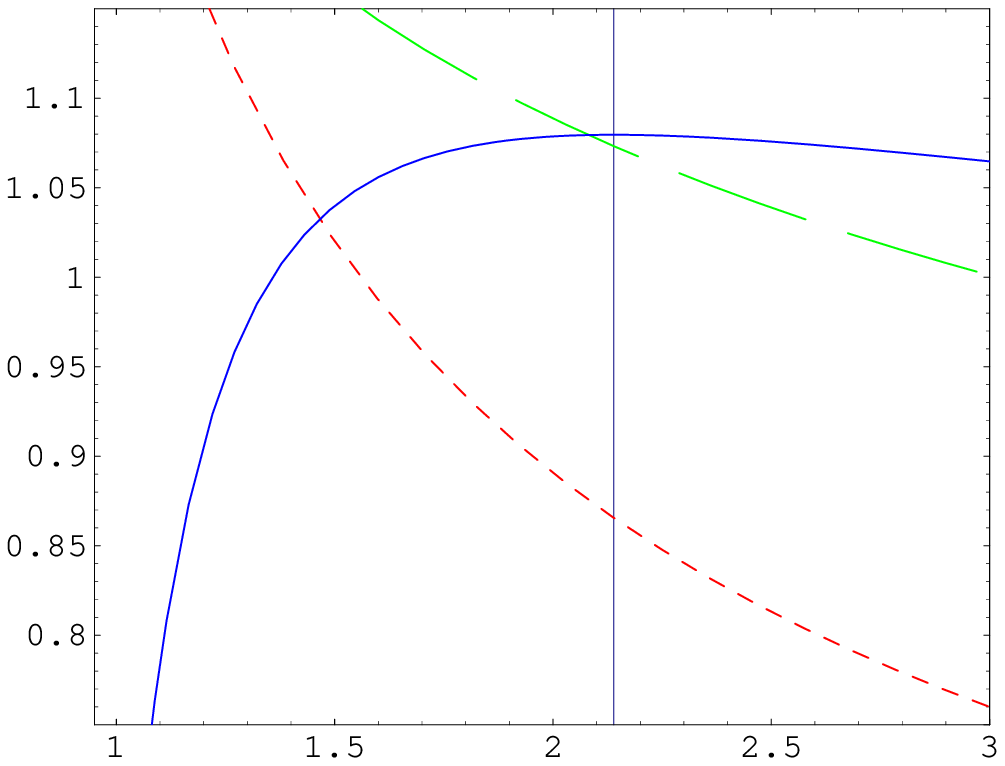}}
\put(250,0){\epsfxsize=7.2truecm\epsffile{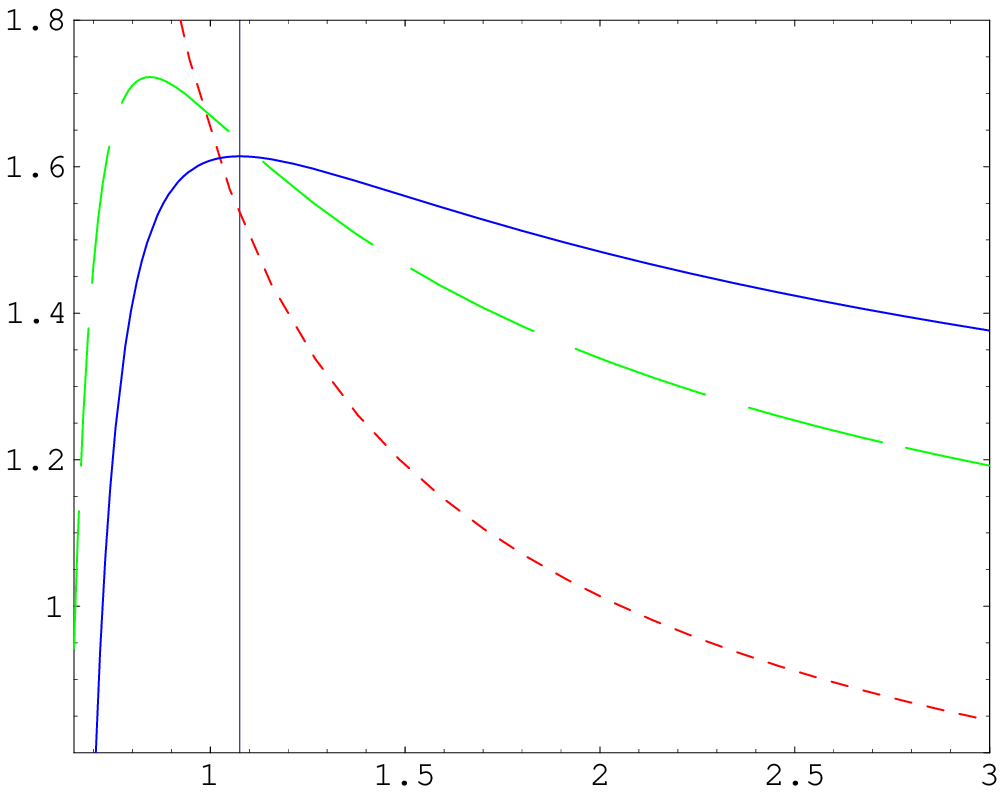}}
\put(10,-190){\epsfxsize=7.2truecm\epsffile{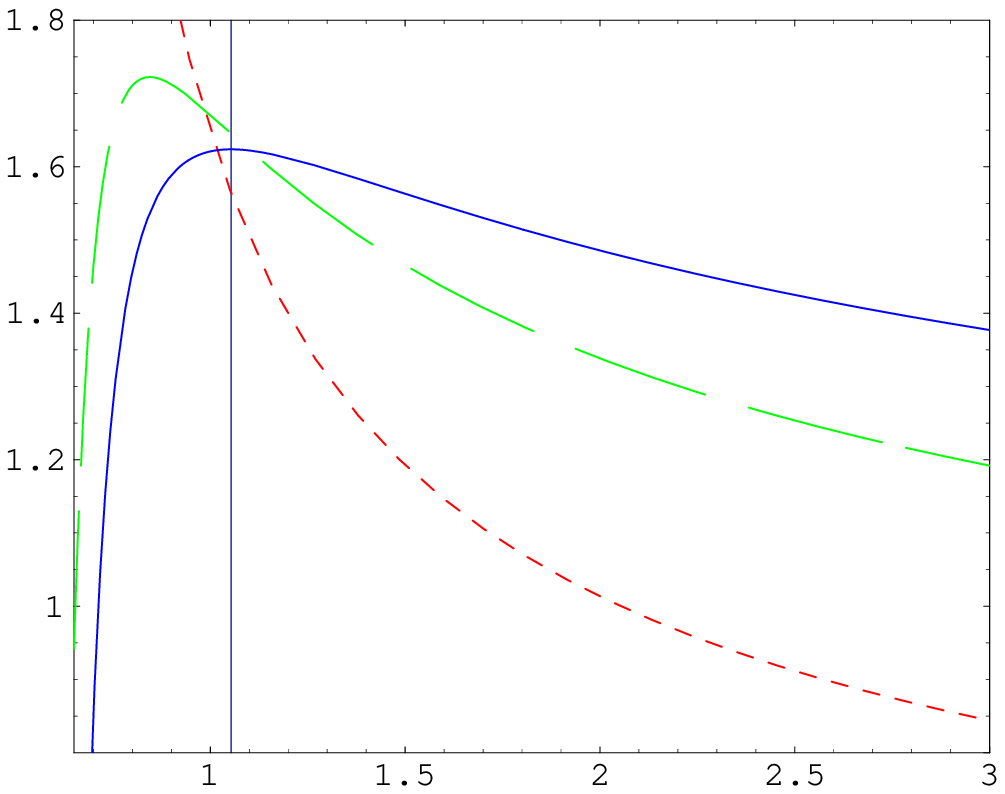}}
\put(250,-190){\epsfxsize=7.2truecm\epsffile{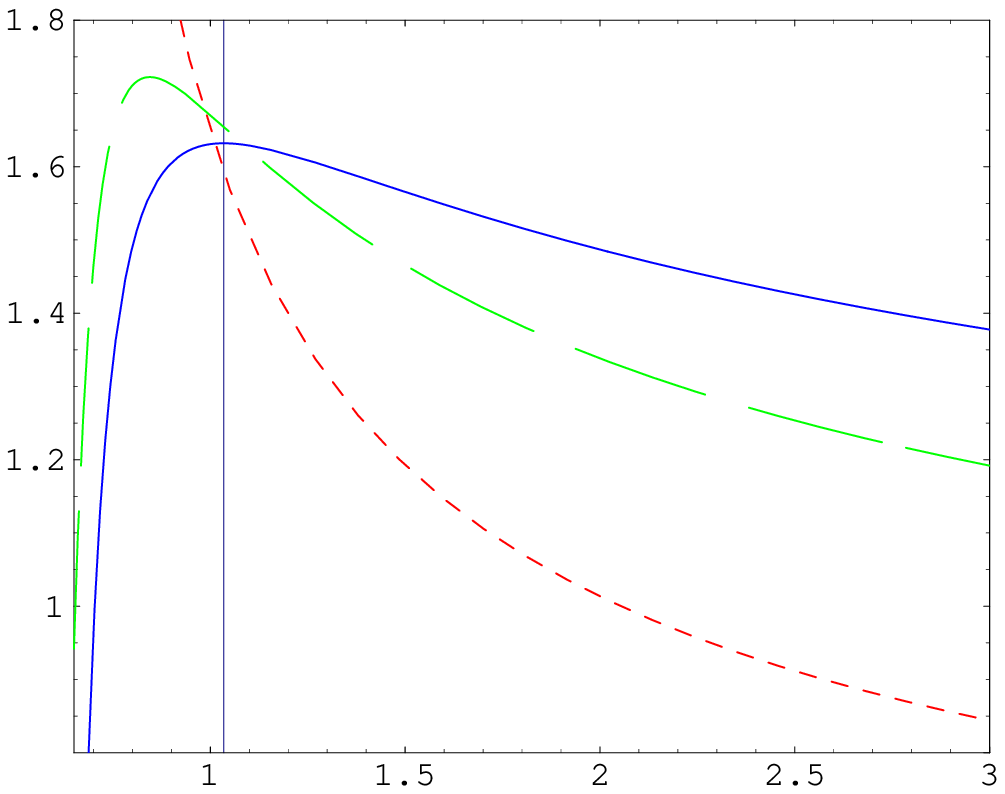}}
\put(180,140){$1S_1$}
\put(420,140){$1P_0$}
\put(180,-50){$1P_1$}
\put(420,-50){$1P_2$}
\put(200,-7){$\mu$}
\put(440,-7){$\mu$}
\put(200,-197){$\mu$}
\put(440,-197){$\mu$}
\put(130,-7){$\mu_X^A$}
\put(295,-7){$\mu_X^A$}
\put(53,-197){$\mu_X^A$}
\put(292,-197){$\mu_X^A$}
\vspace{5mm}
\caption{\footnotesize\it $E_{b\bar{b}}^{(1)}$  (dotted line), 
$E_{b\bar{b}}^{(1)} + E_{b\bar{b}}^{(2)}$ (dashed line) and $
E_{b\bar{b}}^{(1)} + E_{b\bar{b}}^{(2)} + E_{b\bar{b}}^{(3)}$ (continuous line), 
as defined in Eq. (\ref{Emcbis}) 
($(\delta m_b)^{(1,2)}_{m_c} \simeq (\delta m_b)^{(1,2)}_{m_c\to0}$), 
versus $\mu$  for the states  $1S_1$, $1P_0$, $1P_1$ and $1P_2$. 
The parameters are $\mbar_b= 4.190$ GeV, 
$\mbar_c = 1.237$ GeV and  $\Lambda_\msbar^{(3)} = 0.333$ GeV; 
$\mu_X^A$ obtained from the minimal sensitivity prescription, 
Eq. (\ref{scalefixA}), is explicitly shown. The units are GeV.}
\label{figebb1}
\end{figure}

\begin{figure}[\protect{h}]
\makebox[0truecm]{\phantom b}
\put(10,0){\epsfxsize=7.2truecm\epsffile{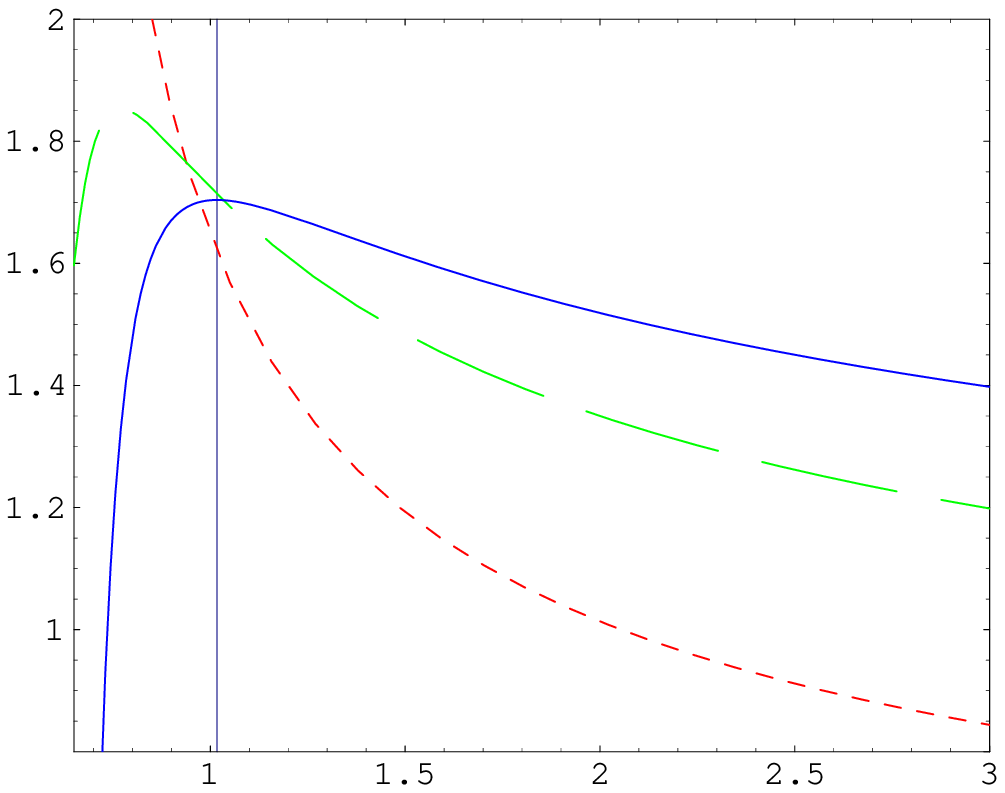}}
\put(250,0){\epsfxsize=7.2truecm\epsffile{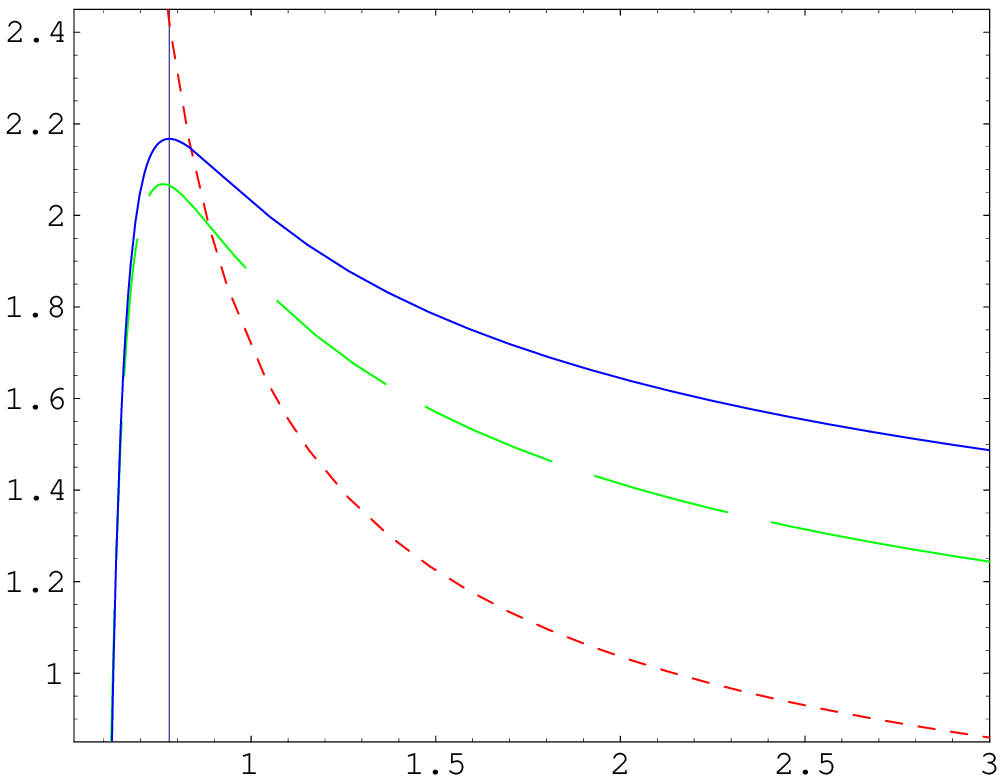}}
\put(10,-190){\epsfxsize=7.2truecm\epsffile{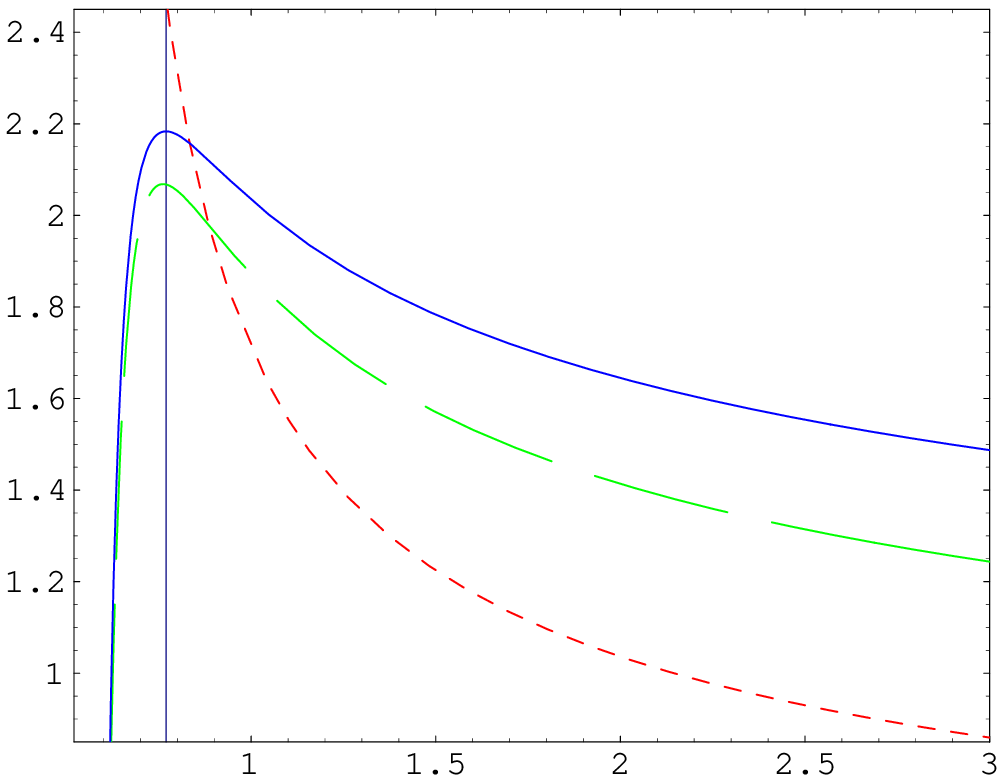}}
\put(250,-190){\epsfxsize=7.2truecm\epsffile{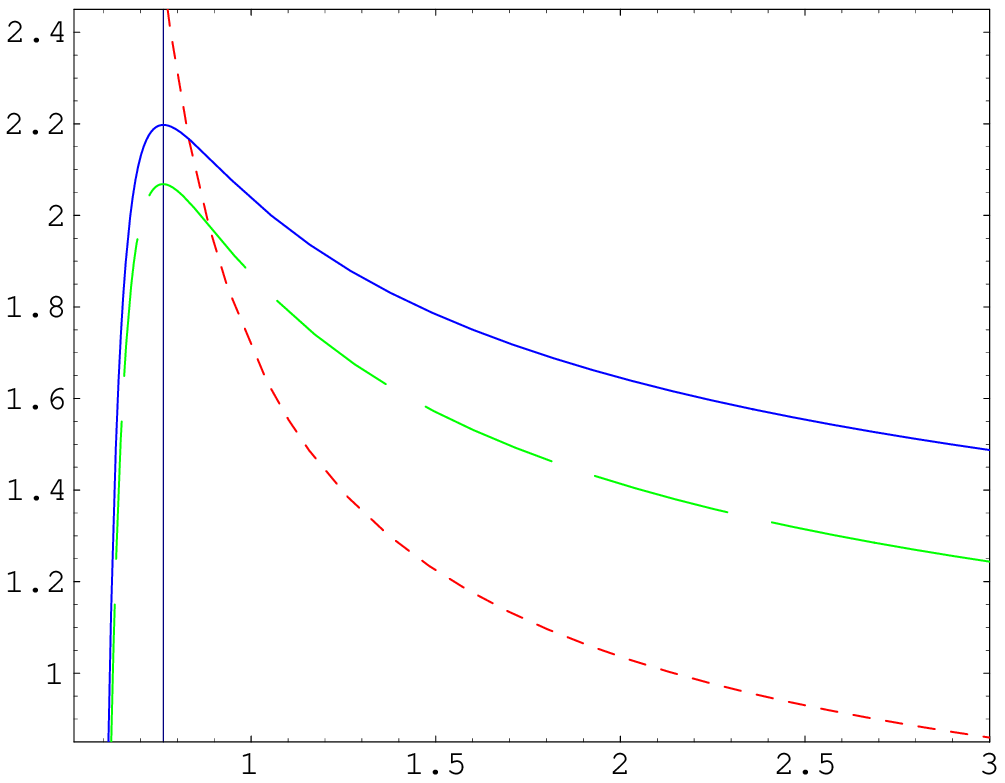}}
\put(10,-380){\epsfxsize=7.2truecm\epsffile{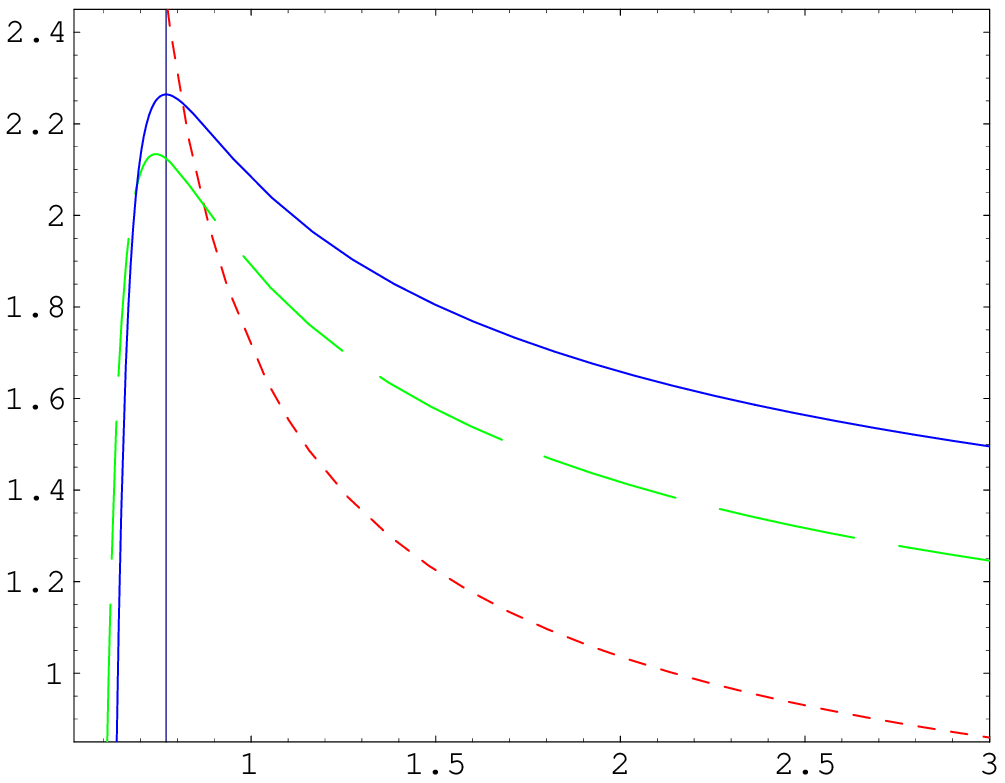}}
\put(250,-380){\epsfxsize=7.2truecm\epsffile{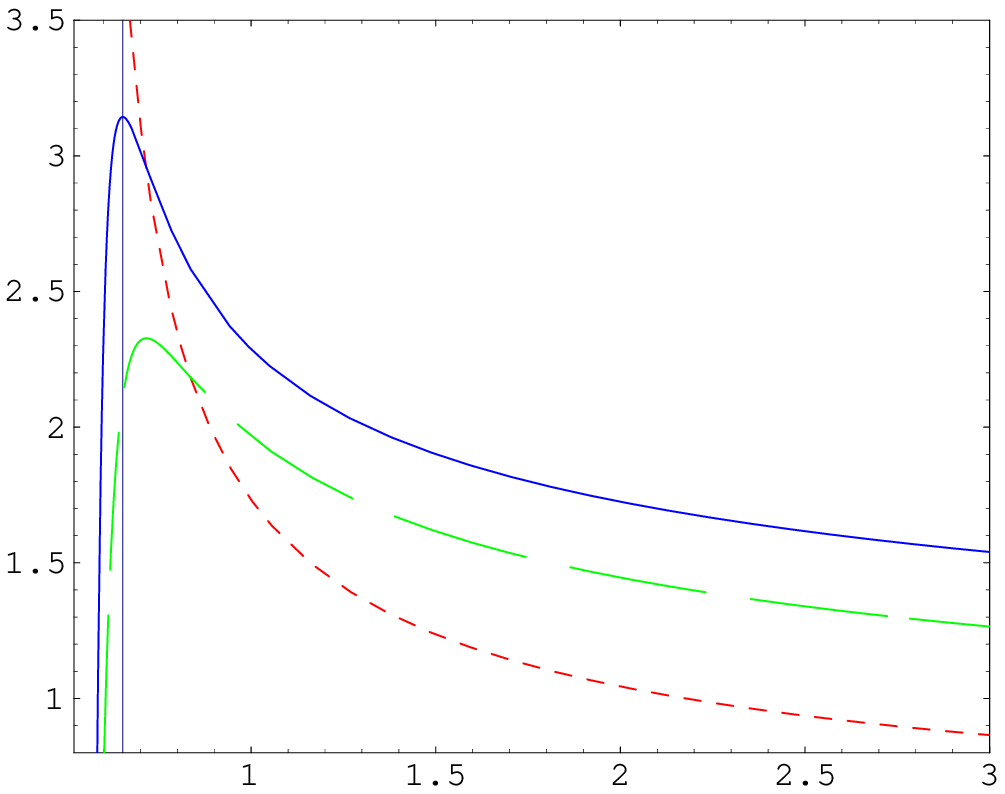}}
\put(180,140){$2S_1$}
\put(420,140){$2P_0$}
\put(180,-50){$2P_1$}
\put(420,-50){$2P_2$}
\put(180,-240){$3S_1$}
\put(420,-240){$4S_1$}
\put(200,-7){$\mu$}
\put(440,-7){$\mu$}
\put(200,-197){$\mu$}
\put(440,-197){$\mu$}
\put(200,-387){$\mu$}
\put(440,-387){$\mu$}
\put(50,-7){$\mu_X^A$}
\put(280,-7){$\mu_X^A$}
\put(40,-197){$\mu_X^A$}
\put(280,-197){$\mu_X^A$}
\put(40,-387){$\mu_X^A$}
\put(270,-387){$\mu_X^A$}
\vspace{5mm}
\caption{\footnotesize\it $E_{b\bar{b}}^{(1)}$  (dotted line), 
$E_{b\bar{b}}^{(1)} + E_{b\bar{b}}^{(2)}$ (dashed line) 
and $E_{b\bar{b}}^{(1)} + E_{b\bar{b}}^{(2)} + E_{b\bar{b}}^{(3)}$ (continuous line), 
as defined in Eq. (\ref{Emcbis}) 
versus $\mu$  for the states  $2S_1$, $2P_0$, $2P_1$, $2P_2$, $3S_1$ and $4S_1$. 
Other notations are the same as in Fig. \ref{figebb1}.}
\label{figebb2}
\end{figure}

\begin{figure}[\protect{h}]
\makebox[0truecm]{\phantom b}
\put(120,0){\epsfxsize=8truecm\epsffile{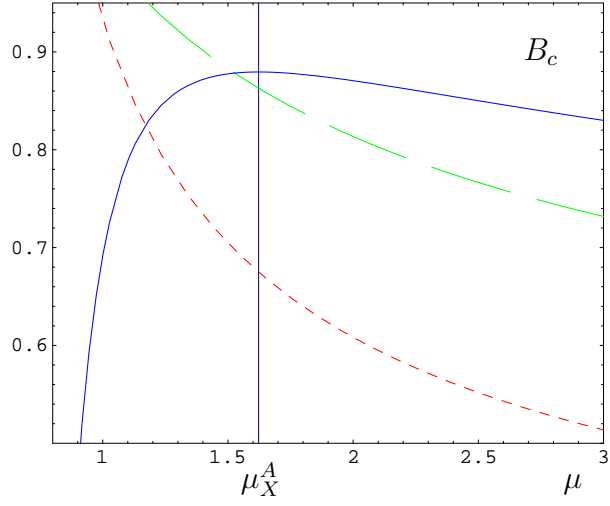}}
\put(315,155){$B_c$}
\put(330,-7){$\mu$}
\put(208,-7){$\mu_X^A$}
\vspace{5mm}
\caption{\footnotesize\it $E_{B_c}^{(1)}$  (dotted line), 
$E_{B_c}^{(1)} + E_{B_c}^{(2)}$ (dashed line) 
and $E_{B_c}^{(1)} + E_{B_c}^{(2)} + E_{B_c}^{(3)}$ (continuous line), 
as defined in Eq. (\ref{EBcbis}) 
versus $\mu$  for the $B_c$. Other notations are the same as in Fig. \ref{figebb1}.}
\label{figebc}
\end{figure}

Using these masses as input, we calculate the energy levels 
of other observed bottomonium and $B_c$ states. In Figs. \ref{figebb1}--\ref{figebc} 
we display the $\mu$ dependence of the different energy levels $E_X$ 
measured from $2\overline{m}_b$ and $\overline{m}_b+\overline{m}_c$, 
respectively (at different orders in 
$\varepsilon$) before scale fixing. The levels, after scale fixing
by Eq. (\ref{scalefixA}), are given in Table \ref{table:spectraA}. 
Theoretical predictions that we consider unreliable, in the sense specified 
in Sec. \ref{secfix}, are not displayed and indicated with a star ($*$). 
Generally, for states that we consider reliably calculable in the perturbative 
approach, the scale dependence decreases as we include more terms of the perturbative series.
For states whose predictions we consider unreliable, the series would become much more convergent 
if we chose a scale different from (typically larger than) $\mu_X$.
The theoretical prediction for $B_c$ is consistent with 
the experimental value, although the experimental error is large.
It is also in agreement with the determination of \cite{bcbv}.

\begin{table}[\protect{h}b]
\begin{center}
\begin{tabular}{c||r|r|r|r|c|c|c}
\hline
State $X$ &$E_X^{\rm exp}~~$ &$E_X~~~$ 
&$E_X^{(1)}$ &$E_X^{(2)}$ &$E_X^{(3)}$ &~~$\mu_X^B$ &$\als^{(3)}(\mu_X^B)$ \\ 
\hline
$\Upsilon(1^3S_1)$  & 9.460 & 9.460~~~~ & 0.876 & 0.204  & 0 & 2.08 & 0.290 \\
$\Upsilon(1 ^3P_0)$ & 9.860 & $9.993^{+75}_{-61}$ & 1.476  & 0.136  & 0 & 1.12 & 0.440 \\
$\Upsilon(1 ^3P_1)$ & 9.893 &$10.002^{+78}_{-63}$ & 1.500  & 0.122  & 0 & 1.10 & 0.448 \\
$\Upsilon(1 ^3P_2)$ & 9.913 &$10.011^{+79}_{-65}$ & 1.521  & 0.109  & 0 & 1.09 & 0.455 \\
$\Upsilon(2 ^3S_1)$ &10.023 &$10.084^{+93}_{-76}$ & 1.593  & 0.111  & 0 & 1.03 & 0.478 \\
$\Upsilon(2 ^3P_0)$ &10.232 & $*$ & $*$ & $*$ & $*$ & $*$ & $*$ \\
$\Upsilon(2 ^3P_1)$ &10.255 & $*$ & $*$ & $*$ & $*$ & $*$ & $*$ \\
$\Upsilon(2 ^3P_2)$ &10.268 & $*$ & $*$ & $*$ & $*$ & $*$ & $*$ \\
$\Upsilon(3 ^3S_1)$ &10.355 & $*$ & $*$ & $*$ & $*$ & $*$ & $*$ \\
$\Upsilon(4 ^3S_1)$ &10.580 & $*$ & $*$ & $*$ & $*$ & $*$ & $*$ \\
$B_c(1^1S_0)$ & $6.4 \pm 0.4$ & $6.309^{+5}_{-3}$ & 0.698 & 0.181 & 0 & 1.53 & 0.347 \\
\hline
\end{tabular}
\end{center}
\vspace{1mm}
\caption{\footnotesize\it
Theoretical predictions for the charmonium, bottomonium and $B_c$ masses, 
when the scales $\mu_X^B$ are fixed through Eq. (\ref{scalefixB}).
The other conventions are the same as in Table \ref{table:spectraA}.
}
\label{table:spectraB}
\end{table}

\noindent {\it Scale-fixing procedure B)}\vspace{2mm}\\
By fixing the bottom and charm $\overline{\rm MS}$ masses on the experimental values of the $\Upsilon(1S)$ 
and $J/\psi$ masses via the scale-fixing procedure of Eq. (\ref{scalefixB}), respectively, we obtain
\bea
\overline{m}_b & = & 4190^{-20}_{+19} ~ {\rm MeV} ,
\label{bmassB}
\\
\overline{m}_c & = & 1241^{-16}_{+14} ~ {\rm MeV} .
\label{cmassB}
\eea
These values are in agreement with the estimates based on the procedure $A)$, given 
in Eqs. (\ref{bmass}) and (\ref{cmass}). The errors refer to the uncertainties 
in $\als^{(5)}(M_Z)$ only. Using these masses as input and fixing the scales 
through Eq. (\ref{scalefixB}), we calculate the energy levels 
of other observed bottomonium and $B_c$ states.
The levels are given in Table \ref{table:spectraB}. 
Reliable theoretical predictions agree well with those given in Table \ref{table:spectraA}. 
\vspace{2mm}\\

Let us compare the present results for the bottomonium states 
with the corresponding results obtained in the $m_c=0$ case in 
Table 2 column (i) of \cite{bsv1}.
The value of the bottom $\overline{\rm MS}$ mass
$\overline{m}_b$ determined from the $\Upsilon(1S)$ mass is
lowered by about 11 MeV by the inclusion of the non-zero charm-mass effects,
for the same input value of $\als^{(5)}(M_Z)$.
The masses of the $n=2$ bottomonium states are raised by about 70--100 MeV
by the charm-mass effects; the masses of the $n=3$ states are raised 
by about 240--280 MeV by these effects.
These tendencies can be understood as originating from an increase
of the binding energies in these systems: if $\mu_X^{A,B} \simlt m_c$, $\als^{(3)}$ 
at the corresponding scale becomes larger than $\als^{(4)}$; 
hence, the higher the energy level is, the more its mass is increased.
For this reason, the value of the input $\als^{(5)}(M_Z)$, which
reproduces the whole level structure of the experimental
data optimally, becomes smaller when we include the charm-mass effects:
in the $m_c=0$ case, $\als^{(5)}(M_Z)=0.1181$ reproduces the
experimental data fairly well, if we take an average of the $S$-wave and $P$-wave 
levels corresponding to each principal quantum number $n$, while
the agreement is better for $\als^{(5)}(M_Z)=0.1161$ when the charm-mass effects are included.
It also remains that the predictions for the $S$--$P$ 
splittings and the fine splittings are smaller than the experimental values,
as noted in \cite{bsv1}. However, it should also be noted that 
for these observables the perturbative expansions considered here 
include only tree-level contributions.
As for the $B_c(1S)$ mass, it is reduced by about 10--20 MeV by the charm-mass 
effects, when $\overline{m}_b$, $\overline{m}_c$ and $\als (M_Z)$ are taken as input.

\subsection{Error estimates} 
\label{ser}
In this section we take the results listed in Table \ref{table:spectraA} as our 
reference calculation and discuss different kinds of uncertainties that may affect them.
Besides non-perturbative corrections, there are four kinds of uncertainties, which we list below.
\begin{table}[h]
\begin{center}
\begin{tabular}{c|l||c|c|c}
\hline
\multicolumn{2}{c||}{} & \multicolumn{3}{c}{Uncertainties of charm-mass effects}\\
\cline{3-5}
\multicolumn{2}{c||}{} & ~~~~~~(i)~~~~~~ & ~~~~~(ii)~~~~~ & ~~~~(iii)~~~~ \\
\hline
\multicolumn{2}{c||}{$\delta \overline{m}_b$} & $+1^{+0}_{-0}$ & $-3^{+1}_{-0}$ & $-2^{+1}_{-0}$ \\
\hline 
& $\Upsilon(1 ^3P_0)$ & $+10^{+2}_{-1}$ & $-15^{-6}_{+3}$ & $-5^{-5}_{+2}$ \\
& $\Upsilon(1 ^3P_1)$ & $+12^{+2}_{-2}$ & $-15^{-7}_{+2}$ & $-4^{-5}_{+1}$ \\
& $\Upsilon(1 ^3P_2)$ & $+14^{+1}_{-2}$ & $-15^{-9}_{+2}$ & $-3^{-6}_{+1}$ \\
& $\Upsilon(2 ^3S_1)$ & $+14^{+2}_{-2}$ & $-18^{-9}_{+3}$ & $-4^{-7}_{+1}$ \\
$\delta E_X$ & $\Upsilon(2 ^3P_0)$ & $+77^{+5}_{+2}$ & $-12^{-29}_{*}$ & $*$ \\ 
& $\Upsilon(2 ^3P_1)$ & $+85^{+5}_{+6}$ & $-4^{-33}_{*}$ & $*$  \\
& $\Upsilon(2 ^3P_2)$ & $+93^{+4}_{+14}$ & $+6^{-39}_{*}$ & $*$ \\
& $\Upsilon(3 ^3S_1)$ & $+79^{+7}_{-1}$ & $-19^{-30}_{+24}$ & $+87^{-35}_{*}$ \\
& $\Upsilon(4 ^3S_1)$ & $*$  & $*$  & $*$  \\
& $B_c(1 ^1S_0)$ & $+1^{-1}_{-1}$ & $-$ & $-$  \\
\hline
\end{tabular}
\end{center}
\caption{\footnotesize\it
Differences between the determination (i) (ii) and (iii)  discussed 
in the item 1) of Sec. \ref{ser} and the theoretical predictions of
Table \ref{table:spectraA}.  The central values correspond to 
$\als^{(5)}(M_Z) = 0.1181$, the upper values to  $\als^{(5)}(M_Z) = 0.1201$, 
the lower values to $\als^{(5)}(M_Z) = 0.1161$. The scale is fixed according to the procedure A) 
of Eq. (\ref{scalefixA}). Those values corresponding to unreliable theoretical predictions 
are indicated with a star ($*$). All dimensionful numbers are in MeV.}
\label{errmc}
\end{table}
\begin{itemize}
\item[1)]{{\it Uncertainties related to the charm-mass effects discussed in this paper.}
Since the charm-mass effects are the main subject of the present paper, their errors
will be discussed in some detail. In evaluating the effects of charm-mass loops in the 
perturbative expansion of the quarkonium levels, we have considered different options 
(see Sec. \ref{seccharm}).  Here we estimate the differences between these options and the 
reference calculation in the previous section. This also shows that the reference 
calculation is the most reliable in the sense discussed at the beginning of Sec. \ref{san}.
In Table \ref{errmc} the following error estimates are listed
(the scales $\mu$ are fixed by Eq. (\ref{scalefixA});
the quoted values correspond to $\als^{(5)}(M_Z) = 0.1181 \pm 0.0020$):
(i) The difference between the results obtained from Eq. (\ref{Emc})
(Eq. (\ref{EBc}) for $B_c$) when the full expression of 
$(\delta m_b)^{(1)}_{m_c}$, as given in Eq. (\ref{m1}),  is used instead of the linear 
approximation $(\delta m_b)^{(1)}_{m_c\to0}$ and the results of Table \ref{table:spectraA}.
(ii) The difference between the results obtained from Eq. (\ref{Emc2})
when $(\delta m_b)^{(1)}_{m_c}$ is approximated by $(\delta
m_b)^{(1)}_{m_c\to0}$ and the results for the bottomonium system of Table \ref{table:spectraA}.
(iii) The difference
between the results obtained from Eq. (\ref{Emc2})
when the full expression of $(\delta m_b)^{(1)}_{m_c}$ is used
and the results for the bottomonium system of Table \ref{table:spectraA}.}
\item[2)]{{\it Uncertainty of $\als^{(5)}(M_Z)$.} This has been taken into account
in the results presented in the previous section and listed in Tables \ref{table:spectraA} 
and \ref{table:spectraB}.
\begin{table}[h]
\begin{center}
\begin{tabular}{c|l||c|c}
\hline
\multicolumn{2}{c||}{} & \multicolumn{2}{c}{Uncertainties from higher-order corrections}\\
\cline{3-4}
\multicolumn{2}{c||}{} &~~~~~~~~~~~(i)~~~~~~~~~~~~ & ~~~~~(ii)~~~~~ \\
\hline
\multicolumn{2}{c||}{$\delta \overline{m}_b$} &  $+2$ &~~~~~~~~~~$\pm 3$  ~~($\pm 25$)\\
\hline 
& $\Upsilon(1 ^3P_0)$ & $-$31 & $\pm 21$ \\
& $\Upsilon(1 ^3P_1)$ & $-$32 & $\pm 22$ \\
& $\Upsilon(1 ^3P_2)$ & $-$33 & $\pm 22$ \\
& $\Upsilon(2 ^3S_1)$ & $-$39 & $\pm 10$ \\
$\delta E_X$ & $\Upsilon(2 ^3P_0)$ & $-$99 & $\pm 102$ \\
& $\Upsilon(2 ^3P_1)$ & $-$100 & $\pm 116$ \\
& $\Upsilon(2 ^3P_2)$ & $-$101 & $\pm 129$ \\
& $\Upsilon(3 ^3S_1)$ & $-$108 & $\pm 140$ \\
& $\Upsilon(4 ^3S_1)$ & $*$ & $*$ \\
& $B_c(1 ^1S_0)$ & $-$3 & $\pm 17$ \\
\hline
\end{tabular}
\end{center}
\caption{\footnotesize\it
Variations of the theoretical predictions of Table \ref{table:spectraA} 
when the uncertainties 3) (i) and (ii) of Sec. \ref{ser} are separately
taken into account. The number in bracket shows a naive estimate of uncertainties
from different $\varepsilon$-expansion schemes.
Those values corresponding to unreliable theoretical predictions are 
indicated with a star ($*$). The input parameter is $\als(M_Z)=0.1181$.
The scale is fixed according to the procedure A) 
of Eq. (\ref{scalefixA}). All dimensionful numbers are in MeV.}
\label{errho}
\end{table}
}
\item[3)]{{\it Uncertainties from higher-order corrections.} 
These have already been considered in \cite{bsv1} 
and may be estimated as follows (see Table \ref{errho}):
(i) 
The difference between the theoretical predictions computed using $\als(\mu)$ 
as obtained by solving the renormalization-group equation numerically 
at 4 loops and the definition of $\als(\mu)$ 
used in the previous sections and specified in Sec. \ref{parameter}.
(ii)
The contributions $\pm|E^{(3)}_X|$ from Table \ref{table:spectraA}
($\pm|E^{(3)}_X|/2$ for $\delta \mbar_b$).
We have not explicitly considered here uncertainties related to different implementations 
of the $\varepsilon$-expansion scheme. Indeed, one could imagine different ways 
of counting in $\varepsilon$ terms that are not directly related 
to the renormalon cancellation in the pole mass and in the static potential,
such as relativistic corrections or spin--orbit and spin--spin interaction terms.
A naive estimate, obtained by subtracting from 
the definition of $E^{(3)}_X$ given in Eq. (\ref{Emcbis}) 
all the terms mentioned above, gives in the $\mbar_b$ case 
$\pm|E^{(3)}_X$(after reshuffling)$|/2 = \pm 25$ MeV. 
However, further investigations are needed on this point.
In \cite{bsv1} also the difference between the theoretical predictions
computed using the 3-loop and the 4-loop running coupling constants 
has been considered. The size of these uncertainties is typically smaller than those listed
here, and, therefore, they are not considered. 
\begin{table}[h]
\begin{center}
\begin{tabular}{c|l||c|c}
\hline
\multicolumn{2}{c||}{} & \multicolumn{2}{c}{Uncertainties in scale fixing}\\
\cline{3-4}
\multicolumn{2}{c||}{} & ~~~~~~(i)~~~~~~ & ~~~~~(ii)~~~~~ \\
\hline
\multicolumn{2}{c||}{$\delta \overline{m}_b$} & 0 & +1 \\
\hline 
& $\Upsilon(1 ^3P_0)$ & $-2$ & $-13$ \\
& $\Upsilon(1 ^3P_1)$ & $-2$ & $-13$ \\
& $\Upsilon(1 ^3P_2)$ & $-1$ & $-14$ \\
& $\Upsilon(2 ^3S_1)$ & 0 & $-23$ \\
$\delta E_X$ & $\Upsilon(2 ^3P_0)$ & $*$ & $-112$\\
& $\Upsilon(2 ^3P_1)$ & $*$ & $-118$\\
& $\Upsilon(2 ^3P_2)$ & $*$ & $-123$\\
& $\Upsilon(3 ^3S_1)$ & $*$ & $-183$\\
& $\Upsilon(4 ^3S_1)$ & $*$ & $*$ \\
& $B_c(1 ^1S_0)$ & +2 &  $-4$ \\
\hline
\end{tabular}
\end{center}
\caption{\footnotesize\it
Variations of the theoretical predictions of Table \ref{table:spectraA} 
when the uncertainties 4) (i) and (ii) of Sec. \ref{ser} are separately
taken into account. Those values corresponding to unreliable 
theoretical predictions are indicated with a star ($*$). 
The input parameter is $\als(M_Z)=0.1181$. All dimensionful numbers are in MeV.}
\label{errscale}
\end{table}
}
\item[4)]{{\it Uncertainties in the scale-fixing procedure.} 
These are also part of the higher-order corrections and 
may be estimated as follows (see Table \ref{errscale}):
(i)
The difference between the theoretical estimates of Table \ref{table:spectraB} 
obtained by fixing $\mu_X=\mu_X^B$ via the condition (\ref{scalefixB})  
and the results of Table \ref{table:spectraA}, 
obtained by fixing $\mu_X=\mu_X^A$ via the condition (\ref{scalefixA}). 
(ii)
For comparison with the error estimates given in conventional analyses,
we consider the (maximal) variations of $\overline{m}_b$ and $E_X$
when we fix the scale as $\pm 10\%$ of the minimal sensitivity scale: 
$\mu = \mu_X^A \times (1\pm 0.1)$, where $\mu_X^A$ is taken from Table \ref{table:spectraA}. 
This is numerically close to the scale-fixing uncertainty evaluation done in \cite{py}.
}
\end{itemize}
A summary of the above uncertainties is given in Table \ref{table:summaryerror}.
The figures are obtained as $\pm$ the absolute value of the maximum 
of the corresponding uncertainties.
In the column labeled ``h.o. corrections'' the maximum is taken from 
the uncertainties of Tables \ref{errho} and \ref{errscale}. 
Note that in Table \ref{errho} we have not calculated the error associated 
with different implementations of the $\varepsilon$-expansion for 
levels higher than the ground state.

\begin{table}[t]
\begin{center}
\begin{tabular}{c|l||c|c|c}
\hline
\cline{3-5}
\multicolumn{2}{c||}{} & ~~~~~~$\Delta \als(M_Z)$~~~~~~ & h.o. corrections &
charm-mass effects \\
\hline
\multicolumn{2}{c||}{$\delta \overline{m}_b$} & $\pm 20$ & $\pm 25$ & $\pm 3$  \\
\hline 
& $\Upsilon(1 ^3P_0)$ & $\pm 75$ & $\pm 31$ & $\pm 15$ \\
& $\Upsilon(1 ^3P_1)$ & $\pm 78$ & $\pm 32$ & $\pm 15$ \\
& $\Upsilon(1 ^3P_2)$ & $\pm 81$ & $\pm 33$ & $\pm 15$ \\
& $\Upsilon(2 ^3S_1)$ & $\pm 93$ & $\pm 39$ & $\pm 18$ \\
$\delta E_X$ & $\Upsilon(2 ^3P_0)$ & $\pm 196$ & $\pm 112$ & $\pm 77$ \\
& $\Upsilon(2 ^3P_1)$ & $\pm 200$ & $\pm 118$ & $\pm 85$ \\
& $\Upsilon(2 ^3P_2)$ & $\pm 203$ & $\pm 129$ & $\pm 93$ \\
& $\Upsilon(3 ^3S_1)$ & $\pm 218$ & $\pm 183$ & $\pm 87$ \\
& $\Upsilon(4 ^3S_1)$ & $*$ & $*$ & $*$ \\
& $B_c(1 ^1S_0)$ & $\pm 4$ & $\pm 17$ & $\pm 1$ \\
\hline
\end{tabular}

\end{center}
\vspace{1mm}
\caption{\footnotesize\it
Summary of the uncertainties discussed in Sec. \ref{ser}. 
The figures are obtained as $\pm$ the absolute value of the maximum 
of the corresponding uncertainties. In the column labeled ``h.o. corrections'' 
the maximum is taken from the uncertainties of Tables \ref{errho} and \ref{errscale}.
All dimensionful numbers are in MeV.}
\label{table:summaryerror}
\end{table}

\begin{table}[h]
\begin{center}
\begin{tabular}{c||c|c|c}
\hline
State $X$ &~~$\mu_X^A$ & Eq. (\ref{ee1}) &  Eq. (\ref{ee2}) \\ 
\hline
$\Upsilon(1 ^3S_1)$ & 2.34 & 0.019 & 0.010 \\
$\Upsilon(1 ^3P_0)$ & 1.08 & 0.059 & 0.025 \\
$\Upsilon(1 ^3P_1)$ & 1.05 & 0.061 & 0.024 \\
$\Upsilon(1 ^3P_2)$ & 1.02 & 0.064 & 0.023 \\
$\Upsilon(2 ^3S_1)$ & 1.00 & 0.070 & 0.035 \\
$\Upsilon(2 ^3P_0)$ & 0.68 & 0.192 & 0.041 \\
$\Upsilon(2 ^3P_1)$ & 0.67 & 0.202 & 0.035 \\
$\Upsilon(2 ^3P_2)$ & 0.65 & 0.226 & 0.017 \\
$\Upsilon(3 ^3S_1)$ & 0.69 & 0.188 & 0.068 \\
$\Upsilon(4 ^3S_1)$ &  *   &   *   &   *   \\
\hline
\end{tabular}
\end{center}
\caption{\footnotesize\it  Pure charm-mass correction of order $\varepsilon^2$
and $\varepsilon^3$ to Eq. (\ref{Emc2}) (last three lines). 
The input parameter is $\als(M_Z)=0.1181$. All quantities are in GeV. }
\label{charmseries}
\end{table}

\subsection{Charm-mass corrections}
In Table \ref{charmseries} we display the pure charm-mass corrections of order $\varepsilon^2$, 
\be
2 (\delta m_b)^{(1)}_{m_c\to0} 
+ (\delta E_{b\bar{b}})^{(1)}_{m_c\to\infty}({\mbar_b},\bar{\rho}), 
\label{ee1}
\ee
and of order  $\varepsilon^3$, 
\be
2 (\delta m_b)^{(2)}_{m_c\to0} 
- {1\over 4} \left({C_F \als^{(4)}(\mu)\over n}\right)^2 (\delta
m_b)^{(1)}_{m_c\to0} 
+ (\delta E_{b\bar{b}})^{(2)}_{m_c\to\infty}({\mbar_b},\bar{\rho}) 
+ {4\over 3}{\als^{(4)}(\mu)\over \pi} (\delta
E_{b\bar{b}})^{(1)}_{m_c\to\infty}({\mbar_b},\bar{\rho}), 
\label{ee2}
\ee
of Eq. (\ref{Emc2}), where the energy is expressed in terms of the 4-flavour coupling. 
The full energy levels can be easily obtained from the data 
displayed in column (ii) of Table \ref{errmc} and from Table \ref{table:spectraA}. 
Table \ref{charmseries} shows that also the disentangled finite charm-mass 
corrections exhibit a reasonable convergence,  
in correspondence to the reliable predictions of Table \ref{table:spectraA}.

\section{Conclusions and Discussion}
\label{scd}
We have computed the bottomonium spectrum within the framework of perturbative QCD, 
including the non-zero charm-mass effects. 
The effects of a charm mass $m_c>0$ as compared to the $m_c=0$ case are to increase 
the level spacings; the effects are larger among the higher levels.
They can be understood as follows: the effective coupling becomes larger at the relevant 
scale when the decoupling of the charm quark is incorporated; consequently, 
the binding energy increases. Since we fixed 
$\overline{m}_b \equiv m_b^{\overline{\rm MS}}(m_b^{\overline{\rm MS}})$ 
on $\Upsilon(1S)$ in our analysis, the net effects are to decrease $\overline{m}_b$ by about 11 MeV 
and to increase the $n=2$ and $n=3$ levels by about 70--100 MeV and 240--280 MeV, respectively.

We have analysed the uncertainties of our predictions within perturbative QCD.
The uncertainties originating from the error of the input $\als^{(5)}(M_Z)$ [Sec. \ref{ser}: 2)]
are as large as (in most cases larger than) other uncertainties, reflecting a high sensitivity of the 
bottomonium spectrum to $\als^{(5)}(M_Z)$. The uncertainties from unknown higher-order corrections 
[Sec. \ref{ser}: 3) and 4)] are consistent with the estimates based on the next-to-leading order
renormalons, i.e. are numerically of the same size as 
$\Lambda_{\rm QCD}\times (a_X \Lambda_{\rm QCD})^2$:
if we approximate $1/a_X \simeq \mu_X^A$, take the values of 
Table \ref{table:spectraA}, and $\Lambda_{\rm QCD} = 300$--$500$ MeV, we obtain 
for the $1S$ state a contribution of order $\pm(5$--$30)$ MeV, for the 
$n=2$ states a contribution of order $\pm(20$--$130)$ MeV and for the $n=3$ states a contribution of 
order $\pm(40$--$220)$ MeV. The uncertainties of the charm-mass effects have
been discussed in Sec. \ref{ser}: 1).

If we take a lower value ($\simeq 0.1161$) of the input $\als^{(5)}(M_Z)$ within its present error, 
the theoretical predictions for the bottomonium spectrum are in agreement with the experimental data 
within the next-to-leading renormalon uncertainties.
Considering the high sensitivity to $\als^{(5)}(M_Z)$, the agreement
(inside the present world-average values) seems to be quite
non-trivial. Thus, the data at our disposal would confirm the analysis done in \cite{bsv1}:
(1) the bulk of the bottomonium spectrum is accessible by perturbative QCD 
up to the $n=3$ states (at least in the scheme of Table \ref{table:spectraA}); 
(2) non-perturbative contributions may be of the type associated 
with the next-to-leading renormalon as estimated above. 

If the true value of $\als^{(5)}(M_Z)$ turns out to
be close to its upper value of $0.1201$, however, independent non-perturbative
effects, which cannot be absorbed into the next-to-leading renormalons,
should exist, with magnitudes of at least about 60 MeV for the $n=2$ states  
and at least about 260 MeV for the $n=3$ states.
In this case, the non-perturbative effects should decrease the higher levels,
resulting in a tendency opposite to that of the effects usually considered by
local gluon condensates or by a linear confining potential.
Unless $\als^{(5)}(M_Z)$ is known more precisely, we cannot
constrain the non-perturbative effects more stringently with our method.
On the other hand, once $\als^{(5)}(M_Z)$ is measured more precisely in the future,
it is certain that we will be able to investigate the non-perturbative
effects on the bottomonium states very much in detail, since there are a number
of energy levels that can be computed reliably in the present perturbative method.

Yet another possible approach is to take advantage of the high sensitivity of the bottomonium 
spectrum to $\als^{(5)}(M_Z)$ and to determine both $\als^{(5)}(M_Z)$ and $\overline{m}_b$ 
from a fit to the experimental data, assuming that non-perturbative effects can be absorbed
into the next-to-leading renormalon uncertainties. Then we may compare the value 
of $\als^{(5)}(M_Z)$ with those obtained from other measurements, for a consistency 
check of the assumption. It is a meaningful method to study the nature of 
the non-perturbative effects.

Therefore, according to our observation that the perturbative predictions, 
within uncertainties, are consistent with the experimental data, we present the value of
the bottom quark $\overline{\rm MS}$ mass as determined from the
$\Upsilon(1S)$ mass in our approach 
\bea
\overline{m}_b = 4190 \pm 20 \pm 25 \pm 3 ~ {\rm MeV},
\label{bmassfinal}
\eea
where the first error refers to the uncertainty in $\als^{(5)}(M_Z)$,
the second error to the higher-order corrections, and the 
third error to the uncertainty of the charm-mass corrections.
It is useful to compare the above result with a recent estimate done in
\cite{rs}, $\mbar_b = 4210 \pm 25 \pm 90$, where the renormalon
cancellation has been implemented in a scheme different from the
$\varepsilon$ scheme adopted here. Taking into account that in \cite{rs}
charm-mass effects have not been included in the calculation (only added to 
the errors) the central value of 4210 MeV is consistent, within errors, with our 
central value of 4190 MeV [or, with the value of Eq. (\ref{bmass0})]. 
The error $\pm 25$ MeV refers to uncertainties in $\als^{(5)}(M_Z)$
and corresponds to our $\pm 20$ MeV. The error $\pm 90$ MeV refers to
charm-mass effects, higher-order perturbative and non-perturbative 
corrections. It corresponds to our errors $\pm 25 \pm 3$ MeV.
Subtracting from $\pm 90$ MeV a $\pm (10$--$20)$ MeV estimate 
of the charm-mass effects that have not been calculated explicitly in \cite{rs}, but
have been included in our analysis, it remains a  $\pm (40$--$50)$ MeV 
difference between the two error estimates. The difference is given by the
size of the non-perturbative corrections. These are guessed to be  $\pm 50$
MeV in \cite{rs}. On the other hand, in the present work, 
the observation that the (perturbative) determinations of the $n=2$ and $n=3$ 
bottomonium levels agree within the present uncertainties with the experimental data 
suggests this to be also the case for the $n=1$ level, i.e. for $\mbar_b$.
In other words, non-perturbative effects would be absorbed into the
present uncertainties.
Hence, higher bottomonium levels provide the additional physical input needed in 
order to constrain the non-perturbative corrections on the $b$ mass.
We also mention that in the analysis of \cite{hoangcm}, 
where $\mbar_b$ has been extracted from the $\Upsilon$ sum rules including 
charm-mass effects, the author obtains  $\mbar_b = 4.17
\pm 0.05$ GeV. This figure is consistent with our estimate.

Let us, finally, remark some general tendencies of the perturbative predictions.
First, evolving $\als^{(n_l)}(\mu)$ from $\als^{(5)}(M_Z)$
using the 3-loop running instead of the 4-loop running, is 
theoretically still consistent within our present accuracy.
In this case, the level spacings of the spectrum become smaller,
since the 4-loop coefficient of the beta function is negative.
Secondly, if we compare the analytic running-coupling
constant (defined in Sec. \ref{parameter} and used throughout this paper)
and the numerical running coupling constant (used in \cite{bsv1}), 
the former is larger at relevant scales.
Hence, the level spacings become smaller if we use the latter coupling.
These variations of the predictions are certainly higher-order uncertainties.
Nevertheless, it is surely worthwhile to further examine the quality of the theoretical
predictions carefully and to seek for most reliable theoretical predictions
at the current reachable accuracy.

\section*{Acknowledgements}
Y.S.\ was supported in part by the Japan-Germany Cooperative Science Promotion Program.
The work of A.V. has been supported by the European Community through the
Marie-Curie fellowship HPMF-CT-2000-00733.
A.V. acknowledges discussions with Antonio Pineda.
\vfill\eject

\appendix
\Appendix{Pole mass expansion coefficients}
\label{appole}
Here, we give the analytic expressions 
for the coefficients $d^{(n_l)}_1$ and $d^{(n_l)}_2$ that appear in Eq. (\ref{pole}). 
The coefficient $d^{(n_l)}_1$ has been calculated in \cite{gray}, 
while the analytic expression of $d^{(n_l)}_2$ can be derived from the result of \cite{polemass2},\footnote{ 
This constant was obtained numerically in \cite{chst} in a certain 
approximation.  
}
the renormalization-group evolutions of $\als^{(n_l)}(\mu)$
and $m_{\overline{\rm MS}}(\mu)$, and the matching condition \cite{lrv}:\footnote{
This relation coincides with Eq. (14) of \cite{polemass2}, which
is given numerically (indirectly through $\beta_0$ 
and $\beta_1$).
Note that,  in the other
formulas of \cite{polemass2}, the coupling of the full theory is used.
}
\begin{eqnarray}
d^{(n_l)}_1 &=&
{\frac{307}{32}} + {\frac{{{\pi }^2}}{3}} 
      + {\frac{{{\pi }^2}\,\ln 2}{9}} - {\frac{\zeta_3}{6}}
+ 
  n_l\,\left( -{\frac{71}{144}} - {\frac{{{\pi }^2}}{18}} \right)
\nonumber \\ &\simeq&
13.4434-1.04137 \, n_l ,
\\ ~~~ \nn\\
d^{(n_l)}_2 &=&
{\frac{8462917}{93312}} + {\frac{652841\,{{\pi }^2}}{38880}} - 
  {\frac{695\,{{\pi }^4}}{7776}} - {\frac{575\,{{\pi }^2}\,\ln 2}{162}} 
\nonumber \\ &&
- 
  {\frac{22\,{{\pi }^2}\,{{\ln^2 2}}}{81}} - 
  {\frac{55\,{{\ln^4 2}}}{162}} 
- 
  {\frac{220\,{\rm Li}_4(\frac{1}{2})}{27}} 
+ {\frac{58\,\zeta_3}{27}} - 
  {\frac{1439\,{{\pi }^2}\,\zeta_3}{432}} + 
  {\frac{1975\,\zeta_5}{216}}
\nonumber \\ &&
+ 
  n_l\,\left( -{\frac{231847}{23328}} - 
     {\frac{991\,{{\pi }^2}}{648}} + {\frac{61\,{{\pi }^4}}{1944}} - 
     {\frac{11\,{{\pi }^2}\,\ln 2}{81}} + 
     {\frac{2\,{{\pi }^2}\,{{\ln^2 2}}}{81}} + 
     {\frac{{{\ln^4 2}}}{81}} 
\right.
\nonumber \\ &&
\left.
+ 
     {\frac{8\,{\rm Li}_4(\frac{1}{2})}{27}} - 
     {\frac{241\,\zeta_3}{72}} \right)  + 
  {n_l^2}\,\left( {\frac{2353}{23328}} + 
     {\frac{13\,{{\pi }^2}}{324}} + {\frac{7\,\zeta_3}{54}}
     \right)  
\nonumber \\ &\simeq &
190.391 - 26.6551\, n_l +  0.652691\, n_l^2 .
\end{eqnarray}

\Appendix{Finite charm-mass effects in the \\ energy expansion}
\label{ape}
In this appendix we give the expression  of $(\delta E_{b\bar{b}})^{(2)}_{m_c\to\infty}(m_b,\rho)$ 
for a $b\bar{b}$ state of quantum numbers $n$, $l$. 
From Eq. (\ref{end}) we get after explicit calculation (we write here explicitly the mass arguments, 
which are the $b$-pole mass and $\rho = 2 n m_{c,{\rm pole}}/({m_{b,{\rm pole}}} C_F \als^{(4)}(\mu))$):
\begin{eqnarray}
& & (\delta E_{b\bar{b}})_{m_c\to\infty} = 
\varepsilon^2 (\delta E_{b\bar{b}})^{(1)}_{m_c\to\infty}(m_{b,{\rm pole}},\rho)
+ \varepsilon^3 (\delta E_{b\bar{b}})^{(2)}_{m_c\to\infty}(m_{b,{\rm pole}},\rho), \\ \nn\\
& & (\delta E_{b\bar{b}})^{(1)}_{m_c\to\infty}(m_{b,{\rm pole}},\rho) = 
{{m_{b,{\rm pole}}} (C_F\als^{(4)}(\mu))^2\over 4n^2}{\als^{(4)}(\mu)\over \pi} \nn\\
& & \hspace{13mm}\times
\left\{{2\over 3} \ln\left({2\over \rho}\right) - {5\over 9} - {2\over 3}
\bigg(\psi(n+l+1)+\gamma_E\bigg) \right\}, \label{e1}\\ \nn\\
& & (\delta E_{b\bar{b}})^{(2)}_{m_c\to\infty}(m_{b,{\rm pole}},\rho) = 
{{m_{b,{\rm pole}}} (C_F\als^{(4)}(\mu))^2\over 4n^2}\left({\als^{(4)}(\mu)\over \pi}\right)^2 \nn\\
& & \hspace{13mm} \times
\left\{ 2 \beta_0^{(4)}\left[ -{1\over 2}\ln \left({2\over \rho}\right)
\left(\ln \left({2\over \rho}\right) + \ln \left({m_{c,{\rm pole}}\over \mu}\right)\right)
\right.\right. \nn\\
& &  \hspace{13mm}
+ \left( {1\over 2} \ln \left({m_{c,{\rm pole}}\over \mu}\right) +\ln \left({2\over \rho}\right)\right)
\left(\psi(n+l+1) + \gamma_E +{1\over 2} \right) \nn\\
& &  \hspace{13mm}
-{2\over 3}\theta(-2+n-l){\Gamma(n-l)\over\Gamma(n+l+1)} 
\sum_{k=0}^{n-l-2} {\Gamma(2l+2+k) \over k! (n-l-k-1)^2}
\nn\\
& & \hspace{13mm}
-{1\over 6}\psi(1+l+n)\left( \psi(1+l+n) -2\right) 
- {1\over 3} {(n-l-1)!\over (n+l)!}\sum_{k=0}^{n-l-2}{(k+2l+1)!\over k! (k+l+1-n)^3} \nn\\
& & \hspace{13mm}
- {n\over 3} {(n+l)!\over (n-l-1)!}\sum_{k=n-l}^{\infty}{k!\over (k+ 2l +1)! (k+l+1-n)^3}
- {\pi^2\over 36} - {1\over 2}\gamma_E^2 - {1\over 2}\gamma_E - {5\over 72}\nn\\
& & \hspace{13mm}
\left.  -{1\over 3}\psi^\prime(n+l+1) - {1\over 3}(\psi(n+l+1))^2 - \psi(n+l+1)\left({5\over 6} 
+ \gamma_E\right) \right] \nn\\
& &  \hspace{13mm}
- {1\over 3}\ln^2 \left({2\over \rho}\right) + \ln \left({2\over \rho}\right) 
\left( -{1\over 2} + {2\over 3} \psi(n+l+1) + {2\over 3}\gamma_E\right) \nn\\
& &  \hspace{13mm}
-{4\over 9}\theta(-2+n-l){\Gamma(n-l)\over\Gamma(n+l+1)} 
\sum_{k=0}^{n-l-2} {\Gamma(2l+2+k) \over k! (n-l-k-1)^2}
\nn\\
& & \hspace{13mm}
-{1\over 9}\psi(1+l+n)\left( \psi(1+l+n) -2\right) 
- {2\over 9} {(n-l-1)!\over (n+l)!}\sum_{k=0}^{n-l-2}{(k+2l+1)!\over k! (k+l+1-n)^3} \nn\\
& & \hspace{13mm}
- {2 n\over 9} {(n+l)!\over (n-l-1)!}\sum_{k=n-l}^{\infty}{k!\over (k+ 2l +1)! (k+l+1-n)^3}
- {\pi^2\over  54} - {13\over 6}\zeta_3 
- {1\over 3}\gamma_E^2 + {1\over 2}\gamma_E \nn\\
& & \hspace{13mm}
\left.
- {183\over 648} -{2\over 9}\psi^\prime(n+l+1) - {2\over 9}(\psi(n+l+1))^2 + \psi(n+l+1)
\left({5\over 18} - {2\over 3}\gamma_E\right) \right\}.
\label{e2}
\end{eqnarray}
Equation (\ref{e1}) corresponds to Eq. (\ref{en1oo}), Eq. (\ref{e2}) has been checked to coincide 
in the $m_c\to\infty$ limit with Eq. (64) of \cite{hoangcm}.

\end{document}